\documentclass[letterpaper,journal]{IEEEtran}
\usepackage{amsmath,amsfonts}
\usepackage{algorithm}
\usepackage{array}
\usepackage{textcomp}
\usepackage{stfloats}
\usepackage{url}
\usepackage{verbatim}
\usepackage{graphicx}
\usepackage{cite}
\usepackage{xcolor} 
\usepackage{hyperref}
\usepackage{multirow}
\usepackage{subcaption}
\usepackage{threeparttable}
\usepackage[commentColor=commentcolor]{algpseudocodex} 
\begin{document}

\bstctlcite{IEEEexample:BSTcontrol}

\title{Open3DBench: Open-Source Benchmark for 3D-IC Backend Implementation and PPA Evaluation}

\author{Yunqi Shi*, Chengrui Gao*, Wanqi Ren, Peng Xie, Siyuan Xu, Ke Xue, Mingxuan Yuan,~\IEEEmembership{Member,~IEEE,} Chao Qian,~\IEEEmembership{Senior Member,~IEEE,} and Zhi-Hua Zhou,~\IEEEmembership{Fellow,~IEEE}
\thanks{Manuscript received April 19, 2021; revised August 16, 2021. (\textit{Corresponding author: Chao Qian.})}
\thanks{Yunqi Shi, Chengrui Gao, Wanqi Ren, Peng Xie, Ke Xue, Chao Qian and Zhi-Hua Zhou are with the National Key Laboratory for Novel Software Technology and the School of Artificial Intelligence, Nanjing University, Nanjing, China (email: qianc@lamda.nju.edu.cn).}
\thanks{Siyuan Xu is with Huawei Noah's Ark Lab, Shenzhen, China.}
\thanks{Mingxuan Yuan is with Huawei Noah's Ark Lab, Hong Kong, SAR.}
\thanks{*Equal contribution.}}

\markboth{Journal of \LaTeX\ Class Files,~Vol.~14, No.~8, August~2021}%
{Shell \MakeLowercase{\textit{et al.}}: A Sample Article Using IEEEtran.cls for IEEE Journals}

\IEEEpubid{0000--0000/00\$00.00~\copyright~2021 IEEE}

\maketitle

\begin{abstract}
This paper introduces Open3DBench, a comprehensive open-source 3D-IC backend benchmark suite built upon the OpenROAD-flow-scripts (ORFS)~\cite{ajayi2019openroad} framework to address the lack of standardized evaluation infrastructures. The framework supports a complete 3D backend design cycle, featuring 3D PDK preparation for tiered stacking, 3D partitioning, placement, timing optimization, 3D routing, and rigorous post-route timing and thermal analysis. For the memory-on-logic (MoL) flow, we implement a graph neural network (GNN)-based 3D partitioning strategy and modular macro placement methods that prioritize regularity and wirelength optimization, respectively. A logic-on-logic (LoL) flow is proposed compatible with ICCAD 2022 and 2023 contest protocols, integrating a reinforcement learning (RL)-based hybrid bonding terminal (HBT) legalization algorithm to resolve manufacturing grid constraints. Experimental results show that our MoL flow achieves substantial PPA improvements over the default 2D ORFS flow, reducing design area by 51.19\%, wirelength by 20.06\%, and negative slack by 42.09\%. For the LoL flow, we provide a comprehensive performance evaluation of top-performing contest-winning binaries. Furthermore, our RL-based legalization reduces HBT displacement by 22.06\% while operating up to 34$\times$ faster than traditional heuristics. Open3DBench is available at \url{https://github.com/lamda-bbo/Open3DBench}, offering a robust, reproducible platform for advancing 3D-IC design automation.
\end{abstract}


\section{Introduction}
\IEEEPARstart{A}s semiconductor process technology approaches its physical limits, scaling within advanced process nodes faces profound obstacles, including escalating EUV lithography costs, exponential design complexity, and diminishing manufacturing yields~\cite{zhao2025physical}. To sustain Moore's Law~\cite{Moore1998cramming}, 3D integrated circuits (ICs) have emerged as a pivotal solution, leveraging vertical stacking to significantly reduce global wirelength, alleviate routing congestion, and enhance timing performance through shortened interconnect paths~\cite{wuu20223dvcache, meta2024isscc}.

Various vertical stacking technologies have been developed to enable inter-die connectivity, yet each presents distinct challenges in terms of density and manufacturability. Traditional through-silicon vias (TSVs) are primarily constrained by their relatively large pitch and the occupation of the device area, which inherently limits integration density~\cite{kim2013study}. While monolithic 3D integration achieves nanometer-scale connectivity through monolithic interior vias (MIVs), it faces formidable barriers regarding ultra-density routing challenges~\cite{panth2014design} and diminished yields. Consequently, face-to-face (F2F) stacking utilizing hybrid bonding terminals (HBTs) has emerged as a pivotal solution that strikes a balance between integration density, manufacturing yield, and cost-effectiveness~\cite{yu2021foundry}. By offering a micron-scale pitch, HBTs satisfy high inter-die bandwidth requirements while facilitating independent die fabrication and wafer-to-wafer bonding, thereby establishing HBT-based F2F stacking as the primary focus of this research.

While 3D integration offers significant performance potential, its practical implementation is hindered by design automation complexities, particularly the exponentially increased search space and intricate physical design rules. A major obstacle for academic research remains the lack of standardized, open-source evaluation infrastructures. In contrast to the mature 2D ecosystem, where open-source tools~\cite{lin2019dreamplace,ajayi2019openroad} provide seamless RTL-to-GDSII flows, the 3D landscape remains fragmented and lacks a unified implementation framework. Consequently, although existing 3D placement algorithms often report improvements in isolated metrics~\cite{liao2024analytical, zhao2025analytical}, the absence of comprehensive power, performance, and area (PPA) analysis prevents rigorous comparative benchmarking. This gap is further widened by the fact that many prominent academic 3D-IC implementation flows~\cite{panth2014shrunk,pentapati2020pin} remain tightly coupled with proprietary commercial engines or in-house scripts. Such dependencies lead to opaque evaluation processes, hindering the consistent PPA benchmarking of 3D-IC methodologies.

To bridge this gap, we present a comprehensive, open-source 3D backend benchmark and implementation framework derived from OpenROAD-flow-scripts (ORFS)~\cite{ajayi2019openroad}. Our framework provides a modular infrastructure to facilitate the integration and rigorous evaluation of diverse 3D physical design algorithms. Our contributions are summarized as follows:

\IEEEpubidadjcol

\begin{itemize}
    \item We introduce \textbf{Open3DBench}, a fully open-source 3D backend flow providing 13 test cases. It supports the complete design cycle, including synthesis, placement, timing optimization, clock tree synthesis, routing, parasitic extraction, and timing and thermal analysis.
    \item We propose a memory-on-logic (MoL) flow featuring a GNN-based partitioner and two macro placement strategies: A tiling-based method for placement regularity and an analytical approach for wirelength optimization.
    \item We integrate a logic-on-logic (LoL) flow compatible with ICCAD 2022 and 2023 contest protocols. This includes an end-to-end PPA evaluation pipeline and a reinforcement learning (RL)-based HBT legalization algorithm to resolve manufacturing grid violations.
    \item We conduct extensive experiments to validate the framework, demonstrating significant PPA improvements and algorithmic efficiency. Our results show that the 3D flow achieves an average area reduction of 51.19\% , wirelength improvement of 20.06\%, and timing enhancement of 42.09\% compared to traditional 2D flows. Furthermore, our RL-based legalization method achieves a 22.06\% reduction in HBT displacement while operating up to 34$\times$ faster than traditional sliding-window heuristics.
\end{itemize}

\section{Related Works}

This section reviews the evolution of 3D-IC backend implementation, categorizing existing works into commercial-tool-based flows and academic solutions.

\subsection{Commercial-Tool-Based Flows}

Despite the emergence of native 3D EDA platforms such as Cadence \textit{Integrity 3D-IC}~\cite{integrity3dic} and Synopsys \textit{3DIC Compiler}~\cite{3diccompiler}, these tools remain in the early stages of maturity. Consequently, leveraging mature 2D commercial engines to realize 3D backends, a strategy often termed ``Pseudo-3D'' flows, remains a dominant approach in academia. This methodology benefits from the rigorous verification and sign-off capabilities of established 2D tools, ensuring industrial viability. We begin by introducing foundational pseudo-3D flows, followed by a discussion of stage-specific optimizations.

\subsubsection{Foundational Pseudo-3D Flows}

Foundational works address the dimensionality mismatch by mapping 3D design spaces onto 2D planes through geometric scaling and parasitic re-characterization. \textbf{Shrunk-2D}~\cite{panth2017shrunk} pioneers the standard cell scaling approach to perform flattened 2D placement, followed by bin-based min-cut partitioning. \textbf{Compact-2D}~\cite{ku2018compact} improves upon this by expanding the placement area to $\sqrt{2} \times \sqrt{2}$ and utilizing RC scaling to emulate 3D parasitics, implementing tier assignment via row-based legalization. Targetting MoL integration, \textbf{Macro-3D}~\cite{bamberg2020macro} introduces abstracted macro models to preserve pin accessibility. Furthermore, \textbf{Pin-3D}~\cite{pentapati2020pin} establishes a ``transparent view'' methodology to enable iterative top-bottom-top timing closure and 3D ECOs.

\subsubsection{Partitioning and Floorplanning}

In the domain of tier partitioning, recent works utilize advanced optimization techniques to address timing and congestion. \textbf{TP-GNN}~\cite{lu2020tpgnn} leverages GNNs to learn high-dimensional node embeddings, capturing hierarchy and timing features to preserve the structural integrity of critical paths. \textbf{Snake-3D}~\cite{huang2025snake} targets the path oscillation issue by employing a gradient-based framework with a differentiable loss function to jointly minimize cutsize, snaking, and local density. Furthermore, the \textbf{PPA-Aware Partitioner}~\cite{jeong2025ppa} adopts a cluster-based integer linear programming (ILP) formulation to concurrently optimize inter-tier cuts, physical cell overlaps, and tier transitions, ensuring rigorous local area balancing. For 3D floorplanning, research focuses on managing hierarchical constraints. \textbf{Cascade2D}~\cite{chang2016cascade2d} and \textbf{Pin-in-the-Middle}~\cite{ku2020pin-in-the-middile} translate vertical alignment constraints into 2D-compatible anchors and fences, enabling standard commercial tools to manage 3D blocks. Advanced methodologies such as \textbf{Hier-3D}~\cite{agnesina2022hier3d} and \textbf{Closing-the-Gap}~\cite{park2025closing} further enhance optimization by introducing sequence-pair representations and congestion-aware gradient floorplanning, respectively.

\subsubsection{Placement and Timing Optimization}

Placement quality is inextricably linked to timing closure, and disjoint optimization in early flows often leads to significant post-partitioning degradation. To mitigate this, \textbf{Snap-3D}~\cite{vanna2021snap} introduces a constrained placement methodology where standard cells are assigned to alternating double-height rows, enabling simultaneous dual-tier optimization within a shared footprint. Recently, machine learning techniques are integrated to guide commercial placers. \textbf{Art-3D}~\cite{murali2022art} applies reinforcement learning to optimize global placement parameters and macro shrinking strategies. Similarly, \textbf{DCO-3D}~\cite{Hsiao2025dco3d} utilizes neural networks to predict 3D routing congestion and optimal cell locations, minimizing the displacement between initial 2D placement and final 3D legalization.

\subsubsection{Routing and Interconnect Optimization}

The final implementation stage focuses on maximizing vertical connectivity and routing resource utilization, encompassing metal layer sharing (MLS) and bonding terminal management. Specifically, \textbf{MLS}~\cite{pentapati2022metal} exploits the continuous routing stack of F2F integration, allowing 2D nets on one die to utilize underutilized routing resources from the adjacent die. \textbf{GNN-MLS}~\cite{hu2025gnnmls} extends this by using GNNs to identify optimal nets for sharing without exacerbating congestion, while \textbf{Closing-the-Gap}~\cite{park2025closing} introduces partial-MLS to selectively share resources in critical regions. Under this F2F integration context, the quality of HBT assignment is also critical for it determines the cross die connectivity. \textbf{Net-to-Pad}~\cite{vanna2025net-to-pad} and \textbf{ViaLegal}~\cite{pentapati2023vialegal} treat HBT assignment as a legalization problem, employing techniques ranging from anchor cell constraints to force-directed algorithms. Taking a holistic approach, \textbf{Unified-Flow}~\cite{kim2025unified} introduces a comprehensive framework, utilizing constraint-driven partitioning to manage HBT count and timing-critical paths, integrated with an RL-assisted clustering scheme for post-routing terminal legalization.

Despite their reliability, these flows suffer from the closed-source nature of commercial tools. The reliance on proprietary scripts and high licensing costs strictly limits reproducibility and accessibility for the broader research community.

\subsection{Academic Solutions}

Open-source implementations are pivotal for democratizing 3D-IC research, offering transparent baselines to overcome the opacity of commercial flows. We categorize these contributions into global placement engines, stage-specific point tool optimizations, and integrated implementation frameworks.

\subsubsection{TSV-Aware 3D Placement}

Early academic research primarily addresses the significant area overhead and stress constraints imposed by TSVs. These works generally fall into partitioning-based flows or analytical placement engines that explicitly manage TSV density. \textbf{T3Place}~\cite{cong2007t3place} utilizes a folding technique to transform wirelength-optimized 2D layouts into 3D stacks, refining layer assignments via a relaxed conflict-net graph to minimize TSV count. Transitioning to continuous optimization, \textbf{Cong et al.}~\cite{cong2010analytical} and \textbf{NTUPlace3-3D}~\cite{hsu2013ntuplace3-3d} introduce force-directed algorithms and weighted-average wirelength models. These solvers manage TSV usage through density penalty functions, treating TSVs as movable objects with significant area. \textbf{ePlace-3D}~\cite{lu2016eplace3d} further advances this domain by incorporating an electrostatics-based density model to handle mixed-size designs. To overcome the limitations of analytical solvers in handling discrete tier assignments, \textbf{HyPlace-3D}~\cite{lin2023hyplace-3d} utilizes a spatial transformation technique to map 3D coordinates onto a 2D continuous space. Integrated with simulated annealing, this approach restores inter-die mobility, enabling dynamic tier reassignment for direct 3D wirelength optimization.

\subsubsection{Hybrid Bonding-Based 3D Placement}

The advent of high-density F2F bonding has shifted the research focus from constraint-driven TSV minimization to the scalability of true-3D placement and HBT co-optimization. Unlike TSV-based flows, these approaches leverage vertical interconnects with no silicon area overhead, enabling aggressive vertical integration. Leveraging GPU acceleration, \textbf{Bistratal-Wirelength}~\cite{liao2024bistratal} and its gradient-optimized successors~\cite{liao2025ultrafast} perform global placement by modeling separate dies within a unified electrostatic system, utilizing divergence theorems to accelerate density computation. To address mixed-size challenges in F2F designs, \textbf{iPL-3D}~\cite{zhao2023ipl3d} and \textbf{Chen et al.}~\cite{chen2024mixed} employ iterative refinement and bipartite matching to co-optimize standard cells and HBTs. More recently, \textbf{Zhao et al.}~\cite{zhao2025analytical} propose an adaptive framework utilizing ILP for macro orientation and switching between 2.5D and true-3D strategies based on design characteristics.

\subsubsection{Partitioning and Legalization}

To facilitate timing closure during the partitioning stage, \textbf{TA-3D}~\cite{kim2025ta3d} employs a balanced K-means clustering algorithm combined with Hungarian matching~\cite{martello1987hungarian}. This approach prioritizes timing constraints by explicitly optimizing the tier assignment of high-delay cells to minimize critical path delays. Addressing post-placement refinement, \textbf{3D-Flow}~\cite{zhao20253dflow} formulates legalization as a minimum cost flow problem on a 3D grid, concurrently resolving cell overlaps and minimizing displacement.

\subsubsection{Routing, Interconnect Optimization, and Verification}

The final implementation stages focus on ensuring robust routability, vertical connectivity, and design rule compliance. \textbf{H3D}~\cite{zhao2025h3d} constructs a heterogeneous grid graph to jointly model routing tracks and HBT resources. It minimizes HBT usage via dynamic programming and breadth-first search (BFS)-based refinement. \textbf{Liu et al.}~\cite{liu2024hbt} utilize quadratic-tree partitioning and minimum spanning tree estimation to guide cost-effective terminal assignment, ensuring scalability for large-scale designs. Finally, to streamline physical verification, \textbf{Chang et al.}~\cite{chang20253ddrc} introduce a unified 3D design rule checking (DRC) flow based on voxel data processing, effectively eliminating the computational inefficiency associated with serial 2D-3D verification steps.

\subsubsection{Integrated Implementation Frameworks}

While point tools demonstrate algorithmic innovation, comprehensive open-source flows remain scarce. \textbf{Open3DFlow}~\cite{zhu2025open3dflow-journal,zhu2025open3dflow-conference} represents the state-of-the-art infrastructure built upon the OpenROAD~\cite{ajayi2019openroad} ecosystem. It executes a ``partition-then-implement'' methodology where dies are synthesized and routed independently, supporting TSV-aware timing, thermal, and signal integrity analysis. However, its architectural reliance on die-by-die implementation precludes the exploration of native 3D optimization techniques, such as cross-die metal layer sharing and concurrent placement. This limitation underscores the critical need for a flexible, native 3D benchmark suite capable of evaluating emerging holistic algorithms.

\section{Overview of Open3DBench}

To advance native-3D design, we propose Open3DBench, a versatile 3D backend framework built upon ORFS~\cite{ajayi2019openroad} but tailored specifically for benchmarking flexibility. Open3DBench introduces a memory-on-logic (MoL) flow (Section~\ref{sec:mol}) enabled by learning-based partitioning and regularity-aware analytical placement, alongside a general PPA evaluation framework for benchmarking potential MoL placement methods. Additionally, it provides a logic-on-logic (LoL) flow (Section~\ref{sec:lol}) compatible with ICCAD 2022~\cite{hu2022benchmark} and 2023~\cite{hu2023benchmark} contest protocols, featuring new benchmarks verified by the contest golden evaluator and supporting full-backend routing, timing, and thermal analysis. Finally, targeting industrial manufacturing constraints, we integrate a state-of-the-art RL-based algorithm~\cite{ren2026reinforcement} for HBT legalization to preserve placement fidelity.

\section{Memory-on-Logic Flow}
\label{sec:mol}
The MoL architecture constitutes a paramount paradigm in 3D IC design, characterized by the vertical stacking of memory dies directly atop a logic die. In this configuration, the bottom tier typically accommodates logic cells, while the top tiers are dedicated exclusively to memory arrays~\cite{chen20203dsram,mutlu2025processingmemory}. By minimizing the physical separation between computation and storage, this vertical integration substantially mitigates data transfer latency and power consumption~\cite{yang20243dstacked}. Moreover, high-density vertical interconnects enable superior memory bandwidth, a critical requisite for data-intensive applications such as deep learning accelerators and high-performance processors. The efficacy of this architecture is evidenced by commercial implementations, including AMD's 3D V-Cache~\cite{wuu20223dvcache} and Sony's stacked CMOS image sensors~\cite{haruta20173layer}. As fabrication and packaging technologies mature, 3D stacking provides a robust roadmap for achieving high-density inter-die connectivity.

\begin{figure}[!t]
\centering
\includegraphics[width=0.46\textwidth]{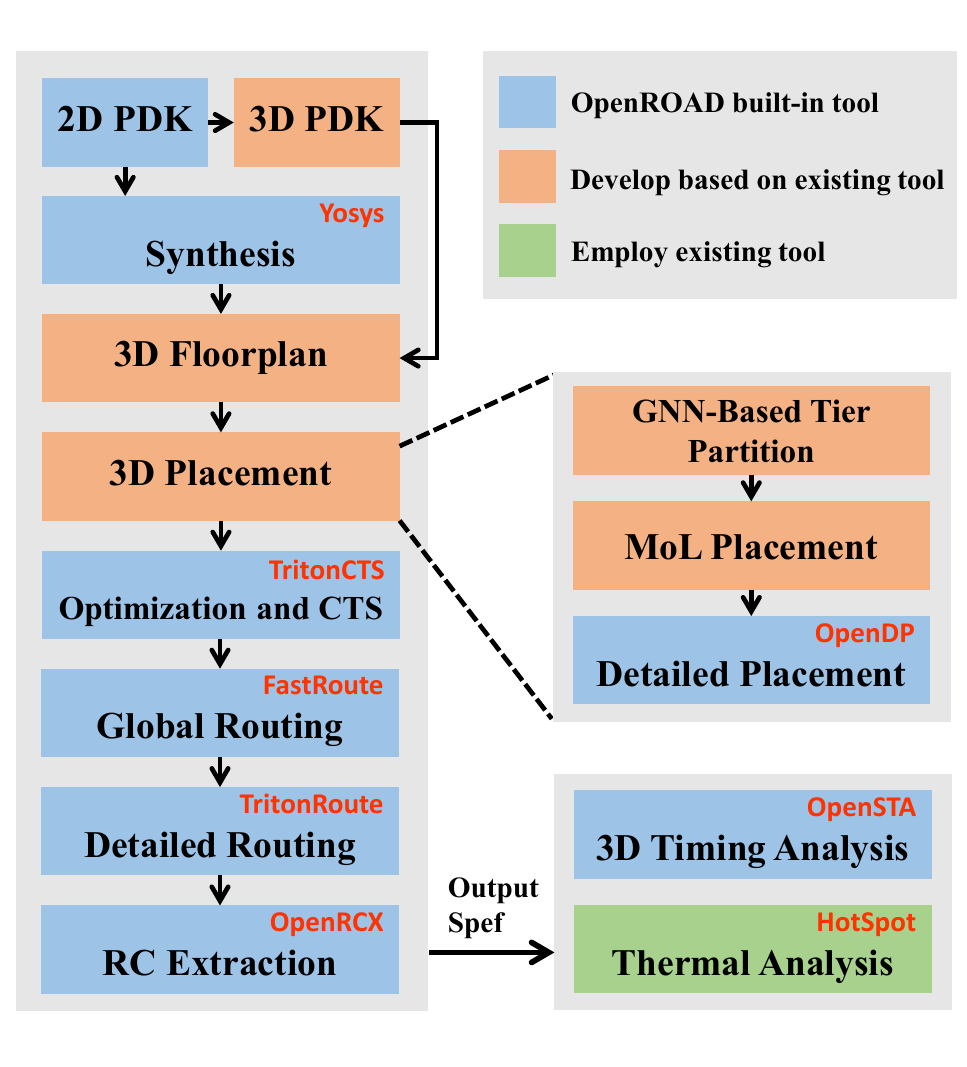}
\caption{Overview of the proposed \textbf{memory-on-logic} flow.}
\label{fig:mem_on_logic_flow}
\end{figure}

Open3DBench implements a comprehensive memory-on-logic design flow, as depicted in Fig.~\ref{fig:mem_on_logic_flow}. The methodology commences with the establishment of an open-source 3D process design kit (PDK) \texttt{NanGate45\_3D}, followed by netlist generation and floorplanning. To address the unique geometric constraints of this architecture, we introduce a GNN-based tier partitioning method, together with an analytical placement algorithm specialized for MoL integration. The flow proceeds with clock tree synthesis (CTS) and timing optimization, concluding with routing and parasitic extraction to facilitate rigorous timing and thermal analyses.

\subsection{PDK Preparation}

We extend the standard 2D \texttt{NanGate45} PDK~\cite{nangate45} to create a 3D-compatible variant, denoted as \texttt{NanGate45\_3D}, by adapting the technology files to capture 3D physical attributes. To model 3D routing resources within a conventional 2D framework, the \textit{techlef} file duplicates the original ten metal layers and inverts them to represent the upper stack. This yields a stacked configuration wherein \texttt{metal1}--\texttt{metal10} are assigned to the bottom die, while \texttt{metal11}--\texttt{metal20} correspond to the top die. Inter-die connectivity is established via an intermediate HBT layer, defined with dimensions of $1 \mu m \times 1 \mu m$ and a pitch of $2 \mu m$, which is defined as a cut layer connecting \texttt{metal10} and \texttt{metal11}.

The \textit{lef} files are updated to specify tier-dependent geometric and pin configurations. To differentiate tier assignments, library components are duplicated and appended with distinct suffixes (e.g., \texttt{buffer\_X16} is bifurcated into \texttt{buffer\_X16\_top} and \texttt{buffer\_X16\_bottom}). Components allocated to the top die undergo pin and routing obstruction remapping to the upper metal layers, whereas bottom-die components retain their original layer definitions.

\subsection{Synthesis and Floorplan}

We employ Yosys~\cite{wolf2013yosys} to generate an initial 2D netlist using the standard 2D PDK. Notably, the transition to a 3D netlist occurs during the subsequent tier partitioning phase. During floorplanning, I/O ports are restricted to the bottom die. Given the reduced footprint of the 3D assembly, we relax pin placement constraints to mitigate the increased I/O density, ensuring feasible port distribution within the limited boundary.

\subsection{GNN-Based MoL Partition}

In a strict memory-on-logic partitioning scheme, memory blocks are restricted to the top die and logic cells to the bottom. However, since the total macro area typically exceeds that of standard cells, this rigidity causes severe tier utilization imbalance, degrading downstream placement and routing quality. To address this, we propose a GNN-based partitioning strategy that selectively migrates macros to the logic die, thereby optimizing area balance and mitigating utilization bottlenecks.

\subsubsection{Problem Formulation and GNN}

We define the tier-partition graph with multiple macro nodes and a single summary-cell node, where edge weights are computed as the number of nets connecting each pair of nodes. Each node is characterized by features including area, width, height, the number of connected cells, and the number of pins. The summary-cell node aggregates all standard cells, with its area derived from the total cell area normalized by the target density and its dimensions determined by the average aspect ratio of all constituent cells. To obtain node embeddings for downstream partition decisions, we employ a weighted graph isomorphism network (GIN)~\cite{xu2018powerful} on this graph. Specifically, let $\mathcal{G}=(\mathcal{V},\mathcal{E})$ denote the tier-partition graph, where each node $v\in\mathcal{V}$ corresponds to a macro (or the summary-cell node), and each edge $(u,v)\in\mathcal{E}$ indicates connectivity. We denote the neighbor set of $v$ as $\mathcal{N}(v)$, the edge weight as $w_{uv}\in\mathbb{R}_{\ge 0}$ (e.g., the number of connecting nets), and the node embedding at layer $k$ as $\mathbf{h}_v^{(k)}\in\mathbb{R}^{d_k}$. Using an MLP $\phi^{(k)}(\cdot)$ as the update function, the weighted-GIN message passing is formulated as
\begin{equation*}
\mathbf{h}_v^{(k)} = \phi^{(k)}\!\left( (1+\varepsilon^{(k)})\,\mathbf{h}_v^{(k-1)} + \sum_{u\in\mathcal{N}(v)} w_{uv}\,\mathbf{h}_u^{(k-1)} \right).
\end{equation*}
where $\varepsilon^{(k)}$ is a learnable scalar that balances the residual strength versus the weighted neighbor aggregation.

\subsubsection{Unsupervised Contrastive Training}

Let $\mathbf{h}_v\in\mathbb{R}^{d}$ denote the final node embedding of $v$. For each anchor node $v$, positive samples are its neighbors $\mathcal{N}(v)$ and the negatives are non-neighbor nodes $\mathrm{Neg}(v)=\mathcal{V}\setminus(\mathcal{N}(v)\cup\{v\})$. The unsupervised contrastive loss is
\begin{equation*}
\begin{aligned}
\mathcal{L}_{\mathrm{unsup}}
&= -\frac{1}{|\mathcal{V}|}\sum_{v\in\mathcal{V}}
\Bigg[
\sum_{u\in\mathcal{N}(v)} \log\tanh\big(\mathbf{h}_v^{\top}\mathbf{h}_u\big)
\\
&\qquad\qquad + \sum_{n\in\mathrm{Neg}(v)} \log\tanh\big(-\mathbf{h}_v^{\top}\mathbf{h}_n\big)
\Bigg],
\end{aligned}
\end{equation*}
This objective encourages macros with similar connectivity to have close embeddings, while pushing weakly-connected (or disconnected) macros apart.

\subsubsection{Embedding-based Partition} 

We perform a two-stage partitioning procedure based on the learned graph embedding. 

\begin{itemize}
    \item \textbf{Initial partition:} We run $k$-means with $k=2$ on the embeddings of all nodes; macros that fall into the same cluster as the summary-cell node are assigned to the bottom die, while the remaining macros are assigned to the top die.
    
    \item \textbf{Area balancing:} To mitigate tier utilization imbalance, we iteratively migrate macros from the higher-area die to the other die, selecting migration candidates according to their proximity to the summary-cell node in the embedding space. For example, when the bottom die is over-utilized, bottom-die macros with the least proximity to cells are relocated to the top die.
\end{itemize}

\begin{figure}[!t]
\centering
\includegraphics[width=0.48\textwidth]{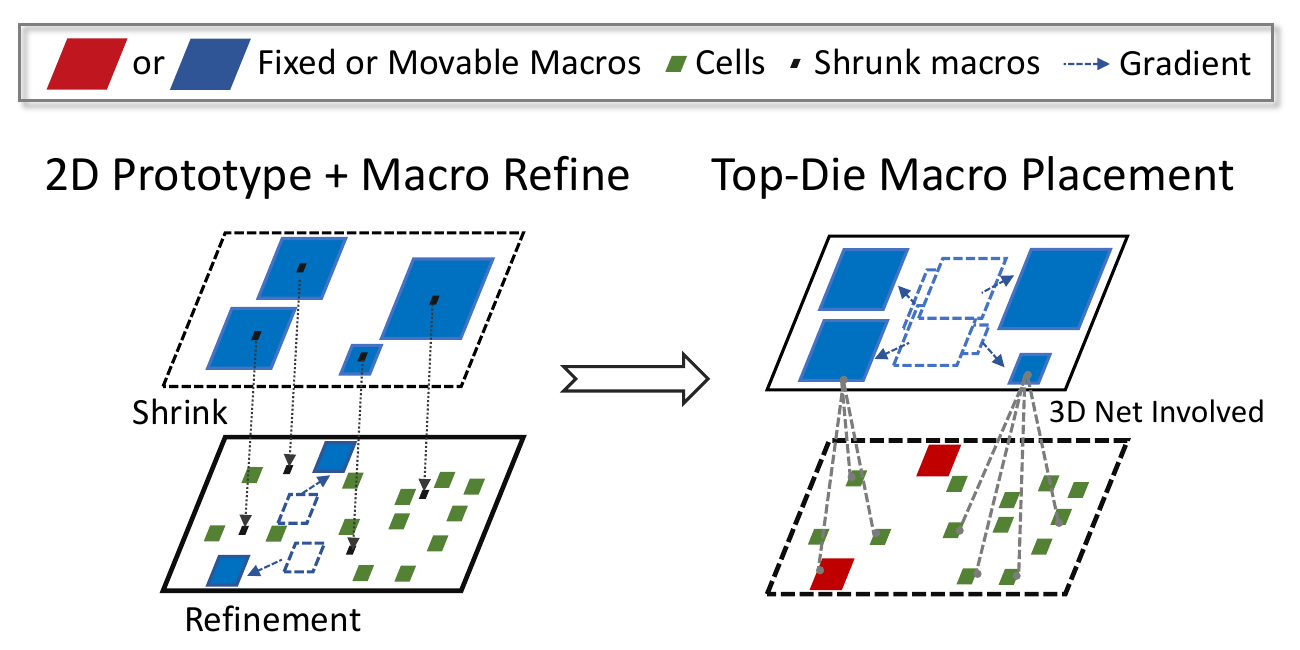}
\caption{Illustration of our analytical pseudo-3D placement process. In each stage, the 2D placement is performed on the 2D plane framed by the solid line.}
\label{fig:place}
\end{figure}

\subsection{MoL Placement}\label{sec:placement}
Following the 3D MoL partition, a 3D placement process is required to optimize the coordinates for both macros and standard cells. To address macro placement systematically, we propose and evaluate two strategies: (1) \textbf{A tiling strategy} that aims to regularly place the macros in an array-like layout. (2) \textbf{An analytical method} that utilizes GPU-accelerated gradient ascent to optimize wirelentgh and regularity. 
\subsubsection{Macro Placement by Tiling} Our tiling implementation employs the skyline packing algorithm~\cite{wei2011skyline} with size-ordered macro sequencing, which places each subsequent macro at the leftmost position of the lowest available skyline. To adequately utilize the memory die, the halo around the macro is adaptively increased to make the highest macro exceed $80\%$ of the die height. After macro positions are fixed, we project the top-die macros to the bottom die with a shrunk size and perform cell placement on the 2D bottom die by DREAMPlace~\cite{lin2019dreamplace}. 

\subsubsection{Analytical Pseudo-3D Placement}
While the tiling strategy provides a structured baseline, we propose a fully analytical, gradient-based pseudo-3D placement framework, where macro coordinates on both dies are optimized via continuous objectives, followed by standard-cell placement on the bottom die. The analytical flow as presented in Fig.~\ref{fig:place} consists of four major steps: 2D prototype, bottom-die macro refinement, top-die macro placement, and bottom-die cell placement. 

\paragraph{2D Prototype} We first construct a 2D prototype to provide a warm start for subsequent optimization. Specifically, all top-die macros are temporarily shrunk to the size of the smallest standard cell and placed together with the remaining instances on the bottom die using DREAMPlace. The resulting 2D layout serves as the prototype.

\paragraph{Bottom-Die Macro Refinement}\label{sec:bottom_die_macro_refinement}
Starting from the prototype, we refine bottom-die macro locations using Nesterov-accelerated gradient descent by minimizing a linear weighted sum of three smoothed objectives: \emph{wirelength}, \emph{peripheral bias}, and \emph{macro overlap}.
Formally, we optimize
\begin{equation*}
\min\; \mathcal{L}_{\mathrm{macro}} = \mathcal{L}_{\mathrm{wl}} + \alpha\,\mathcal{L}_{\mathrm{peri}} + \beta\,\mathcal{L}_{\mathrm{ovlp}},
\end{equation*}
where $\alpha,\beta$ are trade-off weights, $\mathcal{L}_{\mathrm{wl}}=\sum_{e\in\mathcal{N}_{\mathrm{macro}}} \mathrm{WA}_e$ measures macro-involved half-perimeter wirelength~(HPWL), $\mathcal{L}_{\mathrm{peri}}=\sum_{m_i\in\mathcal{M}_{\mathrm{macro}}} P_i$ encourages periphery-aligned placement, and $\mathcal{L}_{\mathrm{ovlp}}$ penalizes macro overlaps.
Here, $\mathcal{N}_{\mathrm{macro}}$ denotes the set of nets incident to at least one bottom-die macro, and $\mathcal{M}_{\mathrm{macro}}$ is the macro set on the bottom die.
\begin{itemize}
    \item \textbf{Wirelength term.} Since the original HPWL is non-smooth, we adopt the weighted-average (WA) wirelength~\cite{hsu2013tsv} as a differentiable surrogate. 
    \item \noindent\textbf{Peripheral bias term.} To encourage boundary-aligned and regular macro placement on the bottom die, we introduce a convex peripheral cost~\cite{pu2024incremacro}. Let $w_{\mathrm{core}}$ ($h_{\mathrm{core}}$) denote the width (height) of the bottom-die core, and let $(x_i,y_i)$ denote the center coordinate of macro $m_i$. The horizontal and vertical peripheral costs are
    \begin{equation*}
        \begin{aligned}
        P_i^{H} &= \left|\frac{w_{\mathrm{core}}}{2}-x_i\right| + \frac{\left(\frac{w_{\mathrm{core}}}{2}\right)^2}{\left|\frac{w_{\mathrm{core}}}{2}-x_i\right|+\epsilon},\\
        P_i^{V} &= \left|\frac{h_{\mathrm{core}}}{2}-y_i\right| + \frac{\left(\frac{h_{\mathrm{core}}}{2}\right)^2}{\left|\frac{h_{\mathrm{core}}}{2}-y_i\right|+\epsilon},\\
        P_i &= P_i^{H} + P_i^{V},
        \end{aligned}
    \end{equation*}
    where $\epsilon>0$ is a small constant for numerical stability. Intuitively, $P_i$ is minimized when each macro $m_i$ is placed close to the boundary.

\item 
\noindent\textbf{Macro overlap term.}
We compute the overlap penalty by multiplying the 1D overlap lengths in $x$ and $y$. Specifically, for each macro pair $(m_i,m_j)$, we first evaluate the 1D overlap length on each axis from their center coordinates and sizes, and then accumulate the product of the two axis-wise overlap lengths across all pairs to form $\mathcal{L}_{\mathrm{ovlp}}$; the exact overlap definition follows~\cite{anonymous2026efficientrefiner}. We apply an exponential schedule for the overlap weight $\beta$ to smoothly transfer from exploration to overlap handling.
\end{itemize}

To ensure macro legality following analytical optimization, we employ a greedy legalization strategy that balances displacement, wirelength, and floorplan regularity. The placement canvas is discretized into a $g \times g$ grid, and macros are prioritized for legalization based on their connectivity density in descending order. For each macro, we evaluate the $K$ nearest unoccupied legal grid sites and select the candidate that minimizes a joint cost function comprising both the total HPWL and the peripheral bias.

\paragraph{Top-Die Macro Placement } We then fix all bottom-die instances and initialize the top-die macros around the center of the top die. Their coordinates are optimized via Nesterov method to minimize a weighted objective $\mathcal{L}_{\mathrm{top}}=\mathcal{L}_{\mathrm{wl}}+\lambda\,\mathcal{L}_{\mathrm{ovlp}}$, where $\mathcal{L}_{\mathrm{wl}}$ is evaluated on nets incident to at least one top-die macro, and $\lambda$ is a trade-off weight that is gradually increased by an exponential schedule. Following this analytical phase, a macro legalization step similar to the procedure in Section~\ref{sec:bottom_die_macro_refinement} is performed. The primary distinction here lies in the selection criteria: For the top die, the cost function is simplified to focus exclusively on minimizing the total HPWL.

\paragraph{Bottom-Die Cell Placement.} Given the legalized macro locations on both dies, we perform standard cell placement and legalization with macros fixed.

Overall, the analytical pseudo-3D placement flow is executed as: \textit{2D prototype} $\rightarrow$ \textit{bottom-die macro refinement} $\rightarrow$ \textit{bottom-die macro legalization} $\rightarrow$ \textit{top-die macro placement} $\rightarrow$ \textit{top-die macro legalization} $\rightarrow$ \textit{bottom-die cell placement}. This analytical pipeline provides a principled, gradient-based placement backbone and yields legal macro positions with explicit HPWL optimization and regularity awareness.

\subsection{Placement Optimization and CTS}

Subsequent to initial placement, post-placement optimization techniques, specifically buffer insertion and gate sizing, are applied to mitigate timing violations and design rule violations (DRVs). Due to the constraints of the MoL architecture, buffer insertion is confined exclusively to the bottom die. To enable this with the standard 2D tool, macros on the top die are abstracted to their minimal physical dimensions (i.e., a single site width and row height) while preserving pin coordinates. This abstraction permits the optimization engine to establish inter-die connectivity without interpreting top-die macros as placement blockages during the bottom-die buffering and sizing processes.

Regarding CTS, the default ORFS~\cite{ajayi2019openroad} configuration is utilized to synthesize the clock network on the bottom die. Clock signals are subsequently propagated to the top-die macros through HBTs.

\subsection{3D Routing with 2D Tools}

By consolidating the interconnect stacks of both dies and the HBT interface into a unified \textit{techlef}, we enable the application of standard 2D routing algorithms to the 3D topology. Global and detailed routing are executed via FastRoute~\cite{pan2006fast} and TritonRoute~\cite{Kahng2021triton}, respectively, with HBTs modeled as standard vertical vias. Although BTAssign~\cite{liu2024hbt} indicates that treating HBTs as vias may result in significant overlaps, our empirical results demonstrate that modern routing engines are robust enough to resolve these constraints natively. By strictly adhering to the defined \textit{techlef}, the router seamlessly accommodates the coarser HBT dimensions ($1 \mu m \times 1 \mu m$ with $1 \mu m$ spacing) alongside the native \texttt{NanGate45} vias ($0.4 \mu m \times 0.4 \mu m$ with $0.88 \mu m$ spacing).

\subsection{Parasitic Extraction}

Parasitic extraction is conducted using OpenRCX~\cite{openrcx}, the native engine within OpenROAD. Precise characterization necessitates rule calibration, as the vertical metal stacking inherent to 3D ICs renders default \texttt{NanGate45} rules insufficient. Consequently, we implement a calibration methodology wherein a commercial sign-off tool generates a reference \textit{spef} based on the unified 3D \textit{techlef}. OpenRCX then derives extraction rules from this golden reference; these calibrated rules are subsequently incorporated into the 3D PDK to facilitate accurate post-route parasitic extraction.

\subsection{Timing Analysis and Thermal Simulation}

The extracted \textit{spef} serves as the input for OpenSTA~\cite{opensta} to perform static timing analysis and verify timing closure. Simultaneously, OpenSTA generates the power distribution profiles essential for thermal modeling. Thermal analysis is subsequently performed using HotSpot~7.0~\cite{han2022hotspot,hotspot}. To achieve an optimal balance between computational efficiency and spatial resolution, the layout is discretized into $10\times 10$ grids, with power density aggregated per grid cell for each die. This methodology is consistently applied across both 2D and 3D implementations to ensure a rigorous comparative analysis of thermal performance.

\section{Logic-on-Logic Flow}\label{sec:lol}

Logic-on-logic (LoL) designs enable the simultaneous placement of macros and standard cells across both the top and bottom dies, offering greater flexibility in 3D integration and leveraging ultra-high-density inter-die interconnects. This approach has been successfully implemented in several industrial systems, including Intel Foveros~\cite{ingerly2019foveros} and the Graphcore Bow IPU~\cite{graphcore_bow_ipu}, demonstrating significant performance improvements. Despite these advancements, LoL designs introduce distinct challenges for existing EDA tools, particularly the need for concurrent and analytical 3D placement. The ICCAD 2022~\cite{hu2022benchmark} and 2023~\cite{hu2023benchmark} contests underscored this gap, stimulating further academic investigation. Nevertheless, most existing algorithms remain focused primarily on wirelength minimization, often overlooking holistic PPA considerations as well as thermal management. To address these limitations, we present a comprehensive LoL design flow that maintains compatibility with contest formats while enabling rigorous backend PPA evaluation.

\begin{figure}[!t]
\centering
\includegraphics[width=0.46\textwidth]{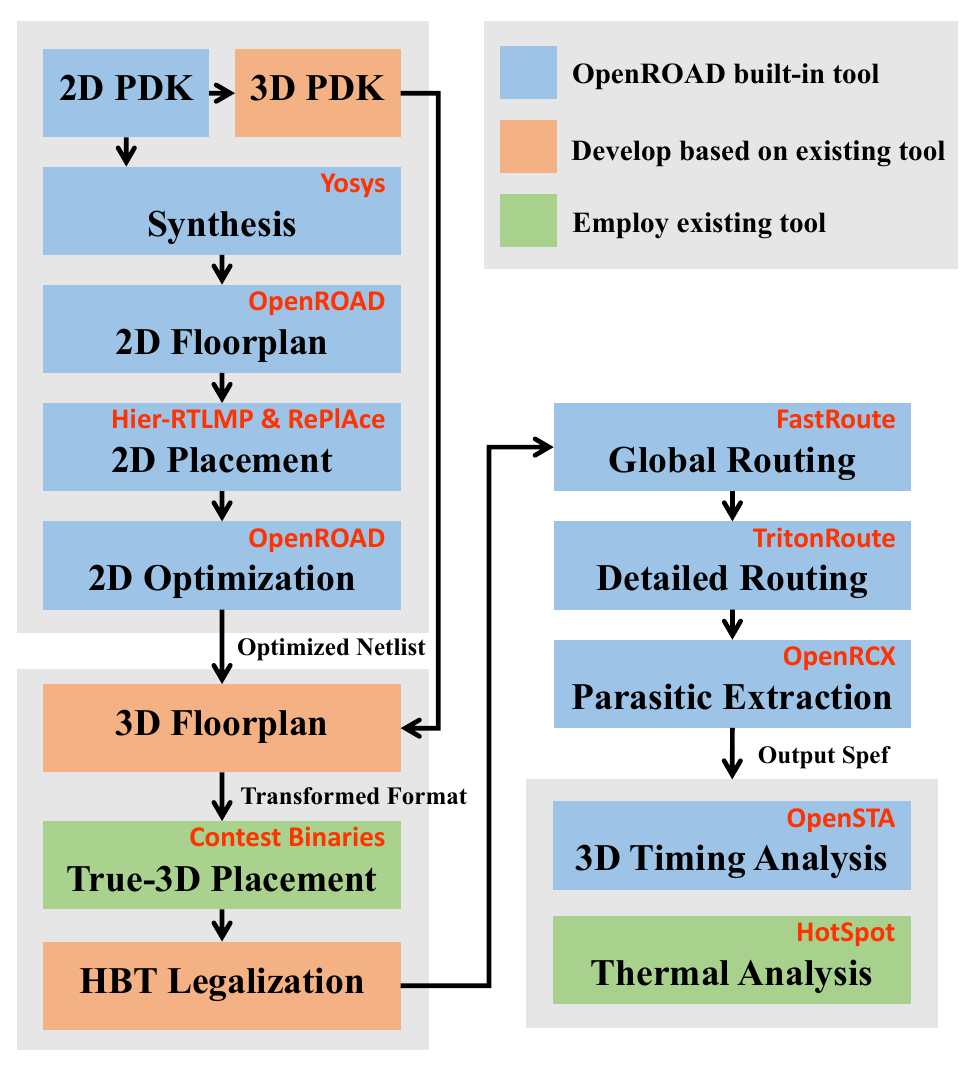}
\caption{Overview of the proposed \textbf{logic-on-logic} flow.}
\label{fig:logic_on_logic_flow}
\end{figure}

The illustration of our proposed LoL flow is given in Fig.~\ref{fig:logic_on_logic_flow}. Compared to the previous MoL flow, our LoL framework introduces critical modifications: it adjusts the placement optimization stage, enables bidirectional translation between \textit{lef}, \textit{def} and contest formats, and integrates explicit algorithms for HBT legalization.

\subsection{PDK Preparation}

PDK modifications generally follow the MoL flow but require specific extensions for HBTs. We model the HBT as a ``virtual buffer'' to facilitate explicit coordinate assignment. Defined with minimal allowable dimensions and negligible parasitics (zero resistance and capacitance), this HBT cell bridges two dies via an input pin on \texttt{metal10} and an output pin on \texttt{metal11}, effectively simulating the electrical transparency of physical bonding interfaces.

\subsection{Netlist and Floorplan Preparation}

Since standard 2D buffering and sizing tools cannot inherently account for cross-die buffering overhead and tier-specific density constraints, we employ an ``optimization-first'' workaround. To this end, we first execute a standard 2D synthesis, placement, and optimization flow to resolve DRVs and stabilize timing, thereby generating an optimized netlist for subsequent 3D placement. Then, we reduce the 2D floorplan area by approximately 50\% and initialize all I/O ports on the bottom tier. Finally, the resulting physical \textit{lef} and \textit{def} files are converted into the standardized contest formats to enable processing by academic 3D placers.

\subsection{True-3D LoL Placement}

To ensure a rigorous evaluation, we utilize the official binaries from the top-performing teams of the ICCAD 2022~\cite{hu2022benchmark} and 2023~\cite{hu2023benchmark} contests to generate initial 3D placement solutions. Specifically, our baseline includes the first and third-place winners from ICCAD 2022 contest, the top three winners from ICCAD 2023 contest, and a recent leading analytical placer~\cite{zhao2025analytical} that outperforms the ICCAD 2023 champion. These advanced placers generally adopt an analytical framework, where density distribution is modeled as a 3D or 2.5D electrostatic field, and wirelength is optimized using a smoothed 3D or 2.5D HPWL loss function.

\subsection{HBT Legalization}

The legalization of HBTs constitutes a critical bottleneck in F2F integration, as the fidelity of HBT placement dictates the subsequent routing quality and overall system performance. While the contest imposes a minimum spacing constraint allowing for a continuous distribution of HBTs, real-world manufacturing processes typically mandate adherence to a fixed grid constraint~\cite{pentapati2023vialegal,liu2024hbt}. Under these constraints, feasible HBT positions are restricted to discrete grid points. Consequently, starting from the ``ideal'' HBT locations derived from the contest winners' global placement, our objective is to map these HBTs to legal grid sites while minimizing total displacement to preserve routing quality and ensure manufacturability.

\subsubsection{Problem Formulation}

\begin{figure}[t!]
\centering
\begin{subfigure}[b]{0.32\columnwidth}
  \centering
  \includegraphics[width=\linewidth, keepaspectratio]{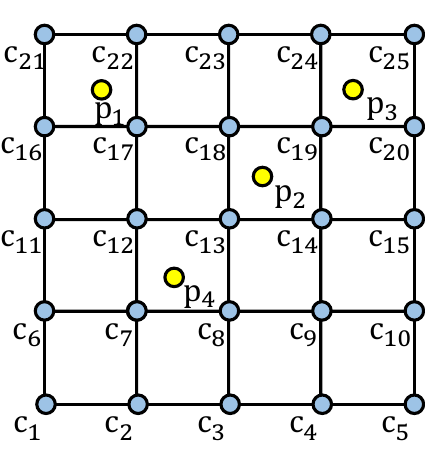}
  \caption{HBT Distribution}
  \label{fig:HBTs}
\end{subfigure}
\hfill
\begin{subfigure}[b]{0.32\columnwidth}
  \centering
  \includegraphics[width=\linewidth, keepaspectratio]{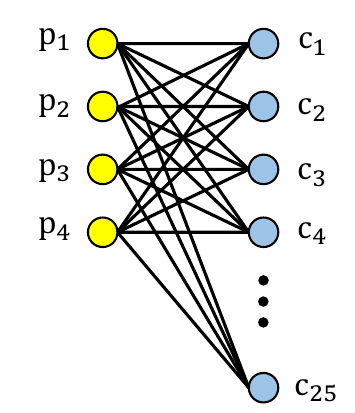}
  \caption{Bipartite Matching}
  \label{fig:matching}
\end{subfigure}
\hfill
\begin{subfigure}[b]{0.32\columnwidth}
  \centering
  \includegraphics[width=\linewidth, keepaspectratio]{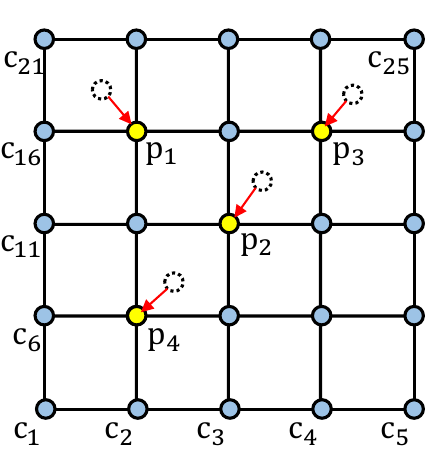}
  \caption{Legalized State}
  \label{fig:legalization}
\end{subfigure}
\caption{Overview of the HBT legalization process: (a) Ideal HBT positions (yellow dots) superimposed on uniform legal candidate sites (blue dots); (b) Formulation as a bipartite matching problem; (c) Finalized legal assignment.}
\label{fig:wbm_problem}
\end{figure}

We formulate the HBT legalization task as a minimum-weight bipartite matching problem, as illustrated in Fig.~\ref{fig:wbm_problem}. Let $\mathcal{P} = \{ \mathbf{p}_1, \dots, \mathbf{p}_n \} \subset \mathbb{R}^2$ denote the set of $n$ ideal HBT positions obtained from the initial placement, and let $\mathcal{C} = \{ \mathbf{c}_1, \dots, \mathbf{c}_m \}$ represent the set of $m$ available legal grid points, where $m \ge n$. The objective is to determine an injective assignment function $f: \mathcal{P} \to \mathcal{C}$ that minimizes the cumulative Manhattan displacement:

\begin{equation}
\min_{f} \sum_{i=1}^{n} \| \mathbf{p}_i - f(\mathbf{p}_i) \|_1
\end{equation}

Although the Hungarian algorithm~\cite{martello1987hungarian} guarantees an optimal solution, its computational complexity of $O(n^3)$ renders it prohibitive for modern industrial-scale designs. Therefore, scalable heuristics are essential to approximate the solution within a reasonable runtime.

\subsubsection{Legalization via Sliding Window}

As a baseline, we adopt the sliding window (SW) strategy proposed in ViaLegal~\cite{pentapati2023vialegal}. This heuristic decomposes the global legalization problem into local sub-problems through a fixed-stride window traversal. Within each window $W$, a local bipartite graph is constructed, and a min-cost max-flow algorithm is employed to resolve the matching.
However, the sequential and uniform nature of this approach presents inherent limitations: A significant portion of the search windows traverse sparse regions lacking resource contention, resulting in computational redundancy, while high-density congestion hotspots may not receive adequate optimization.

\subsubsection{Legalization via Reinforcement Learning}

To address the inefficiencies of fixed traversal strategies, we implement a reinforcement learning (RL)-based framework~\cite{ren2026reinforcement}. This approach replaces the rigid sliding window schedule with an adaptive policy that dynamically prioritizes regions for refinement based on their optimization potential.

\begin{itemize}
    \item \textbf{Initialization and iterative refinement:} The procedure commences with a greedy initialization using spatial hashing to map HBTs to their nearest available candidate grid sites, establishing a valid but potentially suboptimal state $S_0$. We formulate the subsequent refinement as a Markov decision process. Unlike the uniform scan employed in sliding window methods, our RL agent iteratively selects subregions with the highest potential for displacement reduction. At each step $t$, the agent determines the center of a window where the Hungarian algorithm~\cite{martello1987hungarian} is applied to locally re-optimize the assignment. This policy-driven mechanism concentrates computational resources on high-displacement areas, accelerating convergence toward a global optimum.
    
    \item \textbf{State representation and reward design:} We utilize a grid-based multi-channel feature representation to encode the chip state. The state $s_t$ aggregates: (1) \textit{current layout features}, including HBT spatial distribution and local displacement magnitude; (2) \textit{optimization history}, tracking visitation frequency and recent gains to prevent loops; and (3) a \textit{potential map}, which estimates the room for improvement by comparing the current cost against a theoretical lower bound (conflict-free nearest assignment). To guide the agent, we define the reward $R$ as the instant reduction in displacement:
    \begin{equation}
        R(s_t, a_t, s_{t+1}) = \alpha \cdot \left( \text{disp}(f_t) - \text{disp}(f_{t+1}) \right)
    \end{equation}
    where $\text{disp}(f)$ represents the total Manhattan displacement of assignment $f$, and $\alpha$ is a scaling factor.
    
    \item \textbf{Policy training:} A convolutional neural network (CNN) with dual-aspect branches that captures both local features and global context serves as the policy network. The policy is trained on hundreds of synthetic assignment problems via proximal policy optimization (PPO), which enables zero-shot transferability on unseen industrial designs without requiring additional fine-tuning.
\end{itemize}

\subsubsection{Post-Legalization Adjustment}

Following legalization, the design database is updated with the valid HBT coordinates. To enforce 3D routing compliance, we decompose each 3D net into two 2D nets, connecting the respective instance pins to the top and bottom landing pads of the legalized HBT buffer. This adjustment ensures that the router strictly adheres to the intended 3D connectivity.

\subsection{CTS, Routing, and Parasitic Extraction}

In LoL architectures, the dense distribution of sequential elements across tiers challenges conventional 2D CTS tools, which lack the vertical awareness required for balanced 3D clock distribution. Consequently, to isolate the efficacy of our placement and legalization algorithms, we assume an ideal clock network within this flow.

The rest routing and parasitic extraction steps are identical as described in the MoL flow.

\section{Experiments}
\label{sec:experiments}

We implement the proposed 3D framework utilizing the OpenROAD infrastructure\footnote{\url{https://github.com/The-OpenROAD-Project/OpenROAD/commit/fbca14c}}, incorporating DREAMPlace 4.1.0~\cite{chen2023stronger}\footnote{\url{https://github.com/limbo018/DREAMPlace/releases/tag/4.1.0}} to facilitate our proposed placement engine. All evaluations are performed on a Linux workstation powered by a 52-core Intel Xeon CPU (2.60 GHz) and an NVIDIA RTX 2080S GPU, supported by 128 GB of RAM.

To demonstrate the versatility of our framework, we evaluate two mainstream 3D integration paradigms: Memory-on-logic (MoL) and logic-on-logic (LoL). Our benchmark suite comprises thirteen RTL designs sourced from the OpenROAD-flow-scripts (ORFS) repository\footnote{\url{https://github.com/The-OpenROAD-Project/OpenROAD-flow-scripts}}, all synthesized using the \texttt{NanGate45} library~\cite{nangate45}. Specifically, eight designs are employed for MoL and macro-inclusive LoL scenarios, compatible with the ICCAD 2023 contest~\cite{hu2023benchmark} protocols. For macro-free LoL evaluations, we adopt five additional designs aligned with the ICCAD 2022 contest~\cite{hu2022benchmark} protocols. The detailed statistics of these benchmarks are summarized in Table~\ref{tab:statistics}.

\begin{table}[t]
\centering
\renewcommand{\arraystretch}{1.5} 
\caption{Detailed statistics of thirteen designs. Freq. represents the clock frequency of the design.}
\scalebox{1.0}{
\begin{tabular}{|c|c|c|c|c|c|}
\hline
\textbf{Designs}                 & \textbf{\# Macro}   & \textbf{\# Net}                & \textbf{\# Cell}                      & \textbf{Freq. (MHz)}         \\ \hline
\texttt{ariane133} & 132 & 189K & 158K & 250.0 \\ \hline
\texttt{ariane136} & 136 & 194K & 163K & 333.3 \\ \hline
\texttt{black\_parrot} & 24 & 339K & 298K & 500.0 \\ \hline
\texttt{bp\_be} & 10 & 61K & 50K & 384.6 \\ \hline
\texttt{bp\_fe} & 11 & 39K & 32K & 1000.0 \\ \hline
\texttt{bp\_multi} & 26 & 168K & 146K & 1250.0 \\ \hline
\texttt{swerv\_wrapper} & 28 & 118K & 102K & 500.0 \\ \hline
\texttt{bp\_quad} & 220 & 1332K & 1080K & 384.6 \\ \hline
\texttt{aes} & - & 16K & 15K & 1219.5 \\ \hline
\texttt{dynamic\_node} & - & 17K & 14K & 166.7 \\ \hline
\texttt{ibex} & - & 19K & 17K & 454.5 \\ \hline
\texttt{jpeg} & - & 84K & 67K & 1000.0 \\ \hline
\texttt{swerv} & - & 105K & 90K & 500.0 \\ \hline
\end{tabular}}
\label{tab:statistics}
\end{table}

\subsection{Evaluation for MoL Integration}

\begin{table*}[ht!]
\renewrobustcmd{\bfseries}{\fontseries{b}\selectfont}
\renewrobustcmd{\boldmath}{}
\newrobustcmd{\B}{\bfseries}
\caption{Comparative post-route PPA evaluation of 2D and MoL placement strategies across eight benchmarks. \textbf{Bold} values denote the best performance in each category.}
\label{table:main}
\centering
\setlength{\tabcolsep}{5pt} 
\renewcommand{\arraystretch}{1.1}
\begin{tabular}{|c|c|c|c|c|c|c|c|c|c|c|}
\hline
\multirow{2}{*}{Designs} & \multirow{2}{*}{Methods} & \multicolumn{1}{c|}{Area} & \multicolumn{1}{c|}{HPWL} & \multicolumn{1}{c|}{Runtime} & \multicolumn{1}{c|}{GRT\_WL} & \multicolumn{1}{c|}{DRT\_WL} & \multicolumn{1}{c|}{WNS} & \multicolumn{1}{c|}{TNS} & \multicolumn{1}{c|}{Power} & \multicolumn{1}{c|}{T$_{max}$}\\ 
 &  & \multicolumn{1}{c|}{(mm$^2$)} & \multicolumn{1}{c|}{(m)} & \multicolumn{1}{c|}{(s)} & \multicolumn{1}{c|}{(m)} & \multicolumn{1}{c|}{(m)} & \multicolumn{1}{c|}{(ns)} & \multicolumn{1}{c|}{(ns)} & \multicolumn{1}{c|}{(W)} & \multicolumn{1}{c|}{(°C)} \\ \hline

\multirow{3}{*}{\texttt{ariane133}} & \textit{2D-HierRTL} & 2.25 & 4.49 & 667 & 7.75 & 6.51 & -1.45 & -3397 & \B 0.364 & \B 55.16 \\ 
 & \textit{MoL-Tiling} & \B 1.00 & 4.30 & \B 37 & 7.51 & 6.37 & \B -0.54 & \B -706 & 0.376 & 58.80 \\ 
 & \textit{MoL-Analytical} & \B 1.00 & \B 3.76 & 105 & \B 6.98 & \B 5.85 & -0.83 & -1328 & 0.371 & 59.22 \\ \hline

\multirow{3}{*}{\texttt{ariane136}} & \textit{2D-HierRTL} & 2.25 & 4.79 & 1092 & 8.21 & 6.92 & -2.23 & -5720 & \B 0.473 & \B 58.84 \\ 
 & \textit{MoL-Tiling} & \B 1.00 & 4.30 & \B 38 & 7.98 & 6.82 & \B -1.52 & \B -3487 & 0.489 & 62.44 \\ 
 & \textit{MoL-Analytical} & \B 1.00 & \B 3.77 & 105 & \B 7.33 & \B 6.19 & -1.70 & -3904 & 0.483 & 63.25 \\ \hline

\multirow{3}{*}{\texttt{bp}} & \textit{2D-HierRTL} & 1.76 & 5.38 & 233 & 10.82 & 9.06 & -5.76 & -4463 & 0.385 & \B 54.12 \\ 
 & \textit{MoL-Tiling} & \B 0.81 & 3.93 & \B 55 & 9.78 & 8.35 & \B -3.78 & \B -238 & \B 0.372 & 62.00 \\ 
 & \textit{MoL-Analytical} & \B 0.81 & \B 3.71 & 108 & \B 9.50 & \B 8.08 & -3.80 & -332 & \B 0.372 & 60.64 \\ \hline

\multirow{3}{*}{\texttt{bp\_be}} & \textit{2D-HierRTL} & 0.56 & 2.40 & 39 & 3.30 & 2.92 & \B -0.95 & \B -133 & 0.152 & \B 51.65 \\ 
 & \textit{MoL-Tiling} & \B 0.30 & \B 1.78 & \B 20 & 2.88 & 2.56 & * & * & 0.147 & 62.25 \\ 
 & \textit{MoL-Analytical} & \B 0.30 & \B 1.78 & 70 & \B 2.69 & \B 2.35 & * & * & \B 0.146 & 59.25 \\ \hline

\multirow{3}{*}{\texttt{bp\_fe}} & \textit{2D-HierRTL} & 0.48 & 1.51 & 32 & 2.02 & 1.79 & -1.37 & -642 & 0.298 & \B 78.38 \\ 
 & \textit{MoL-Tiling} & \B 0.24 & 1.07 & \B 17 & 2.06 & 1.87 & \B -0.99 & \B -343 & 0.292 & 88.08 \\ 
 & \textit{MoL-Analytical} & \B 0.24 & \B 1.03 & 60 & \B 1.94 & \B 1.73 & \B -0.99 & -472 & \B 0.290 & 79.25 \\ \hline

\multirow{3}{*}{\texttt{bp\_multi}} & \textit{2D-HierRTL} & 1.21 & 3.87 & 113 & 5.78 & 4.95 & -7.39 & -10544 & 1.095 & \B 79.05 \\ 
 & \textit{MoL-Tiling} & \B 0.64 & 2.75 & \B 28 & \B 4.59 & \B 3.85 & -5.02 & \B -3489 & 1.015 & 106.43 \\ 
 & \textit{MoL-Analytical} & \B 0.64 & \B 2.66 & 81 & 4.60 & 3.88 & \B -4.98 & -4133 & \B 1.010 & 102.39 \\ \hline

\multirow{3}{*}{\texttt{swerv\_wrapper}} & \textit{2D-HierRTL} & 1.10 & 3.41 & 84 & \B 5.11 & \B 4.36 & -1.75 & -1423 & 0.243 & \B 52.77 \\ 
 & \textit{MoL-Tiling} & \B 0.56 & 3.24 & \B 20 & 5.38 & 4.77 & -0.75 & \B -521 & 0.231 & 62.01 \\ 
 & \textit{MoL-Analytical} & \B 0.56 & \B 2.74 & 83 & 5.14 & 4.50 & \B -0.72 & \B -521 & \B 0.228 & 61.87 \\ \hline

\multirow{3}{*}{\texttt{bp\_quad}} & \textit{2D-HierRTL} & 12.96 & 33.20 & 1349 & 53.22 & 45.15 & -2.69 & -36501 & 1.815 & 65.21 \\ 
 & \textit{MoL-Tiling} & \B 6.25 & 33.10 & \B 265 & 54.47 & 47.90 & -1.21 & -15444 & 1.783 & \B 65.12 \\ 
 & \textit{MoL-Analytical} & \B 6.25 & \B 29.00 & 454 & \B 51.46 & \B 44.95 & \B -0.87 & \B -10072 & \B 1.779 & 66.56 \\ \hline

\multicolumn{2}{|c|}{\multirow{2}{*}{Avg. Impr. \textit{MoL}\textsuperscript{\dag} vs. \textit{2D-HierRTL}}} & \multirow{2}{*}{51.19\%$\textcolor{red}{\downarrow}$} & \multirow{2}{*}{20.06\%$\textcolor{red}{\downarrow}$} & \multirow{2}{*}{46.99\%$\textcolor{red}{\downarrow}$} & \multirow{2}{*}{7.35\%$\textcolor{red}{\downarrow}$} & \multirow{2}{*}{6.21\%$\textcolor{red}{\downarrow}$} & \multirow{2}{*}{42.09\%$\textcolor{red}{\downarrow}$} & \multirow{2}{*}{61.13\%$\textcolor{red}{\downarrow}$} & \multirow{2}{*}{2.35\%$\textcolor{red}{\downarrow}$} & \multirow{2}{*}{-12.73\%$\textcolor{blue}{\uparrow}$}\\
\multicolumn{2}{|c|}{\multirow{2}{*}{}} &  &  &  &  & &  &  & &\\ \hline
\multicolumn{2}{|c|}{\multirow{2}{*}{Avg. Impr. \textit{MoL-Analytical} vs. \textit{MoL-Tiling}}} & \multirow{2}{*}{Equal} & \multirow{2}{*}{8.28\%$\textcolor{red}{\downarrow}$} & \multirow{2}{*}{-191.88\%$\textcolor{blue}{\uparrow}$} & \multirow{2}{*}{5.01\%$\textcolor{red}{\downarrow}$} & \multirow{2}{*}{5.89\%$\textcolor{red}{\downarrow}$} & \multirow{2}{*}{-4.63\%$\textcolor{blue}{\uparrow}$} & \multirow{2}{*}{-22.94\%$\textcolor{blue}{\uparrow}$} & \multirow{2}{*}{0.66\%$\textcolor{red}{\downarrow}$} & \multirow{2}{*}{2.10\%$\textcolor{red}{\downarrow}$}\\
\multicolumn{2}{|c|}{\multirow{2}{*}{}} &  &  &  &  & &  &  & &\\ \hline

\end{tabular}
\begin{tablenotes}
\item *Timing optimization fails to converge due to unresolved maximum capacitance violations, resulting in unphysically large negative slack.
\item \textsuperscript{\dag}The average improvement is computed as the mean improvement of both \textit{MoL-Analytical} and \textit{MoL-Tiling} methods relative to \textit{2D-HierRTL}.
\end{tablenotes}
\end{table*}

To evaluate the efficacy of the proposed MoL flow, we summarize the post-route PPA metrics in Table~\ref{table:main}. We benchmark our two placement strategies, \textit{MoL-Tiling} and \textit{MoL-Analytical}, against the state-of-the-art 2D OpenROAD flow (\textit{2D-HierRTL}), which employs Hier-RTLMP~\cite{kahng2024hier} for macro placement and RePlAce~\cite{cheng2019replace} for cell placement. In Table~\ref{table:main}, ``HPWL" is reported after global placement and legalization; ``Runtime" accounts for the total duration of macro and cell placement; ``GRT\_WL" and ``DRT\_WL" denote global and detailed routed wirelengths, respectively. Timing metrics (``WNS" and ``TNS") are measured post-routing with parasitic extraction, and ``$\text{T}_{max}$" indicates the peak steady-state temperature via HotSpot~\cite{han2022hotspot} simulation.

As shown in the aggregate results (final rows), 3D integration achieves an area reduction of over 50\% through die stacking and an average HPWL reduction of $\sim$20\% by shortening vertical interconnects. Notably, placement runtime is slashed by $\sim$47\%. This acceleration is attributed to our GPU-accelerated framework, which replaces the CPU-intensive simulated annealing and analytical engines used in the 2D baseline with a GPU-native 3D placement pipeline.

\begin{figure}[ht!]
\centering
\includegraphics[width=\linewidth, keepaspectratio]{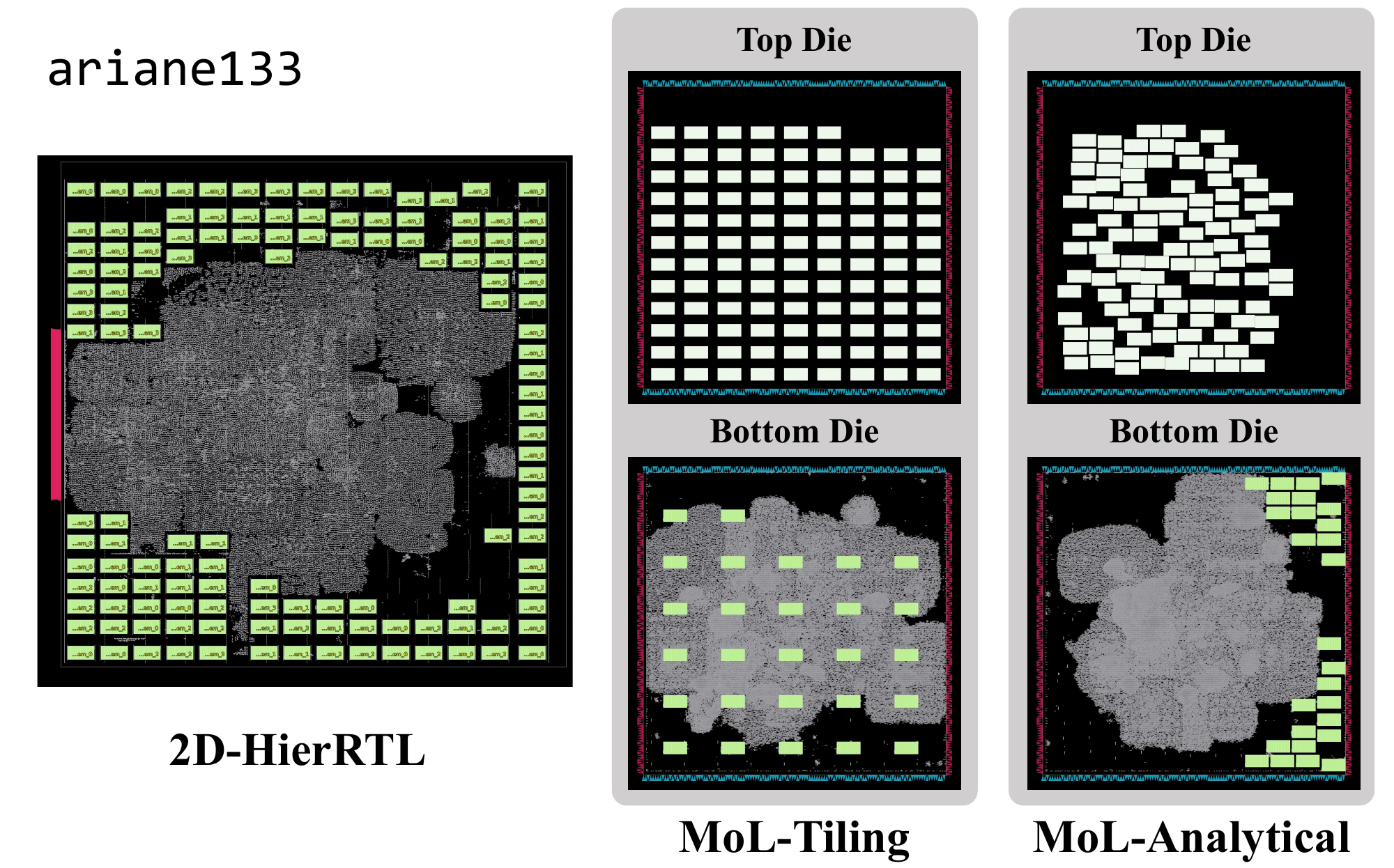}
\caption{Comparative placement topologies for \texttt{ariane133}. While \textit{MoL-Analytical} (right) minimizes inter-tier wirelength through vertical alignment, \textit{MoL-Tiling} (center) maintains superior spatial uniformity, facilitating downstream timing closure compared to the 2D baseline (left).}
\label{fig:mol_visualization}
\end{figure}

While MoL methods consistently improve routed wirelengths (GRT\_WL and DRT\_WL), the gains ($\sim$6-7\%) are less pronounced than those for HPWL. This discrepancy likely arises because the current placement objective does not explicitly reserve cell padding for post-placement optimizations (e.g., buffering and gate sizing). Consequently, the legalization required after these optimizations can induce cell displacement, partially offsetting the initial wirelength benefits. Nevertheless, 3D integration delivers remarkable timing improvements, yielding 42\% and 61\% reductions in WNS and TNS, respectively. Furthermore, it lowers power consumption by reducing net capacitance and the requirement for aggressive cell sizing. Thermal overhead is observed as expected, with T$_{max}$ increasing by 12.7\% on average, which highlights the inherent thermal-density trade-offs in 3D ICs.

\begin{table*}[ht!]
\renewrobustcmd{\bfseries}{\fontseries{b}\selectfont}
\renewrobustcmd{\boldmath}{}
\newrobustcmd{\B}{\bfseries}
\caption{Quantitative comparison of backend performance for LoL integration using benchmarks adapted for the ICCAD 2023 contest binaries. We evaluate the binaries of the contest winners (1st place \textit{cadb0013}~\cite{chen2024mixed}, 2nd place \textit{cadb1038}, and 3rd place \textit{cadb1049}) against a recent state-of-the-art placer \textit{TCAD'25}~\cite{zhao2025analytical}.}
\label{table:iccad2023}
\centering
\renewcommand{\arraystretch}{1.12}
\begin{tabular}{|c|c|c|c|c|c|c|c|c|c|c|c|}
\hline
\multirow{2}{*}{Designs} & \multirow{2}{*}{Methods} & \multirow{2}{*}{Status} & \multicolumn{1}{c|}{Terminal} & \multicolumn{1}{c|}{HPWL} & \multicolumn{1}{c|}{Runtime} & \multicolumn{1}{c|}{GRT\_WL} & \multicolumn{1}{c|}{DRT\_WL} & \multicolumn{1}{c|}{WNS} & \multicolumn{1}{c|}{TNS} & \multicolumn{1}{c|}{Power} & \multicolumn{1}{c|}{T$_{max}$} \\ 
 &  &  & (\#) & (m) & (s) & (m) & (m) & (ns) & (ns) & (W) & (°C) \\ \hline

\multirow{4}{*}{\texttt{ariane133}} 
 & \textit{cadb0013} & Pass & 3767 & 2.73 & 331 & 4.98 & 4.18 & -0.10 & -36 & 0.255 & 56.89 \\ 
 & \textit{cadb1038} & Pass & 3206 & 2.80 & 179 & 5.11 & 4.40 & 0.52 & 0 & 0.265 & 58.81 \\ 
 & \textit{cadb1049} & Pass & 7066 & 2.34 & 1018 & 5.13 & 4.34 & 0.60 & 0 & 0.224 & 56.22 \\
 & \textit{\textit{TCAD'25}}    & Pass & 4934 & 2.56 & 97 & 4.90 & 4.02 & 0.58 & 0 & 0.237 & 55.48 \\ \hline

\multirow{4}{*}{\texttt{ariane136}} 
 & \textit{cadb0013} & DRT Fail & 4235 & 2.77 & 391 & 5.05 & - & - & - & - & - \\ 
 & \textit{cadb1038} & DRT Fail & 3153 & 4.02 & 244 & 6.44 & - & - & - & - & - \\ 
 & \textit{cadb1049} & Pass & 12362 & 2.50 & 1029 & 5.35 & 4.51 & -0.43 & -473 & 0.365 & 62.26 \\
 & \textit{\textit{TCAD'25}}    & Pass & 4793 & 2.69 & 91 & 5.06 & 4.15 & -0.93 & -1796 & 0.281 & 58.38 \\ \hline

\multirow{2}{*}{\texttt{black\_parrot}*} 
 & \textit{cadb0013} & Pass & 12790 & 3.13 & 634 & 6.18 & 5.20 & -3.55 & -8 & 0.411 & 64.19 \\ 
 & \textit{\textit{TCAD'25}}    & Pass & 7667 & 3.47 & 111 & 6.69 & 5.64 & -3.63 & -4 & 0.432 & 64.93 \\ \hline

\multirow{2}{*}{\texttt{bp\_be}} 
 & \textit{cadb0013} & Pass & 1577 & 0.73 & 238 & 2.50 & 2.27 & -0.16 & -5 & 0.117 & 57.65 \\ 
 & \textit{\textit{TCAD'25}}    & Pass & 2702 & 0.72 & 40 & 2.49 & 2.27 & -2.83 & -727 & 0.135 & 60.73 \\ \hline

\multirow{2}{*}{\texttt{bp\_fe}} 
 & \textit{cadb0013} & PL Fail & - & - & - & - & - & - & - & - & - \\
 & \textit{\textit{TCAD'25}}    & Pass & 2790 & 0.80 & 39 & 2.17 & 2.05 & -0.86 & -311 & 0.235 & 76.62 \\ \hline

\multirow{2}{*}{\texttt{bp\_multi}} 
 & \textit{cadb0013} & Pass & 3232 & 1.67 & 611 & 3.48 & 2.95 & -4.96 & -1478 & 0.857 & 95.68 \\ 
 & \textit{\textit{TCAD'25}}    & Pass & 7151 & 1.68 & 87 & 3.59 & 3.02 & -4.96 & -2566 & 0.702 & 84.12 \\ \hline

\multirow{2}{*}{\texttt{swerv\_wrapper}} 
 & \textit{cadb0013} & Pass & 2073 & 1.98 & 496 & 3.93 & 3.43 & -0.26 & -2 & 0.491 & 90.52 \\ 
 & \textit{\textit{TCAD'25}}    & DRT Fail & 5755 & 2.05 & 60 & 3.99 & - & - & - & - & - \\ \hline

\end{tabular}

\begin{tablenotes}
\item *Results for \textit{cadb1038} and \textit{cadb1049} are omitted for the last five test cases due to consistent placement failures (PL Fail).
\end{tablenotes}

\vspace{2em} 

\centering
\includegraphics[width=0.9\linewidth, keepaspectratio]{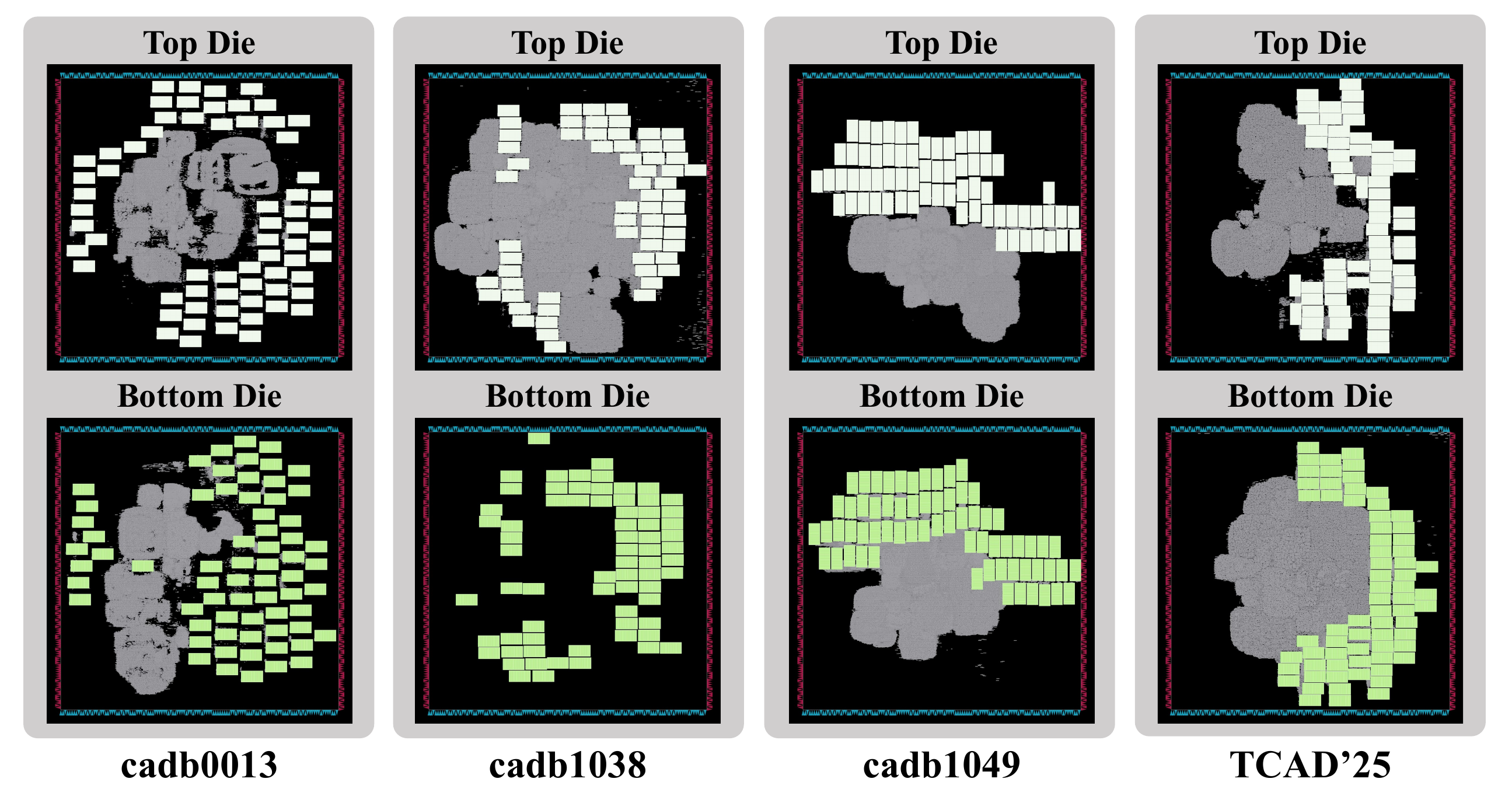}
\captionof{figure}{Comparative placement topologies for the \texttt{ariane133} benchmark in the LoL integration scenario. The layouts illustrate distinct cell distribution strategies between the top three ICCAD 2023 contest winners (\textit{cadb0013}, \textit{cadb1038}, \textit{cadb1049}) and \textit{TCAD'25} across the Top (upper row) and Bottom (lower row) dies.}
\label{fig:lol}

\end{table*}


Comparing the two 3D strategies, \textit{MoL-Analytical} consistently outperforms \textit{MoL-Tiling} in wirelength (8.3\% lower HPWL) due to its concurrent inter-tier optimization. However, this wirelength advantage does not always translate into better timing performance; on average, \textit{MoL-Tiling} achieves slightly superior WNS and TNS. This trade-off is visually elucidated in Fig.~\ref{fig:mol_visualization}: \textit{MoL-Analytical} tends to cluster top-die macros centrally to minimize vertical connections, creating high-density congestion zones on the bottom die that complicate buffer insertion. Conversely, \textit{MoL-Tiling} enforces a more uniform macro distribution, preserving routing resources for timing-critical logic. A notable exception is \texttt{bp\_quad}, where \textit{MoL-Analytical} naturally yields a uniform macro distribution, thereby outperforming \textit{MoL-Tiling} across all metrics. In terms of efficiency, \textit{MoL-Tiling} remains significantly faster as it avoids the iterative overhead inherent in the analytical macro placement.

\subsection{Evaluation for LoL Integration}

\subsubsection{Benchmarking with ICCAD 2023 Contest Winners}

Table~\ref{table:iccad2023} provides a comprehensive backend performance comparison between the ICCAD 2023 contest winners (\textit{cadb0013}~\cite{chen2024mixed}, \textit{cadb1038}, and \textit{cadb1049}) and the state-of-the-art \textit{TCAD'25} placer~\cite{zhao2025analytical}. Following the contest protocol, ``\# of Terminal'' denotes the count of HBTs, which corresponds to the number of 3D nets. ``HPWL'' represents the 3D half-perimeter wirelength validated by the official golden evaluator.

The robustness of each methodology is first evaluated through its completion status. While \textit{TCAD'25} successfully generates feasible placements for all test cases, the contest-winning binaries exhibit varying degrees of fragility. Specifically, \textit{cadb1038} and \textit{cadb1049} encounter frequent placement failures (``PL Fail'') in larger designs. Furthermore, detailed routing failures (``DRT Fail'') are observed across multiple methods. This bottleneck is primarily attributable to the 2D-native nature of TritonRoute; the 3D PDK configuration introduces a non-monotonic routing resource density (declining then increasing across the stack), which disrupts the conventional 2D-centric routing heuristics and impedes convergence. In contrast, \textit{TCAD'25} demonstrates superior algorithmic robustness, achieving full-flow completion in six out of seven designs.

\begin{table*}[ht!]
\renewrobustcmd{\bfseries}{\fontseries{b}\selectfont}
\renewrobustcmd{\boldmath}{}
\newrobustcmd{\B}{\bfseries}
\caption{Impact of cell padding on backend design quality metrics, evaluated using the \textit{TCAD'25}~\cite{zhao2025analytical} placer. The comparison highlights the trade-off between wirelength and routability, with the best valid results marked in \textbf{bold}.}
\label{table:cell_padding}
\centering
\setlength{\tabcolsep}{5pt} 
\renewcommand{\arraystretch}{1.12}
\begin{tabular}{|c|c|c|c|c|c|c|c|c|c|c|c|}
\hline
\multirow{2}{*}{Designs} & \multirow{2}{*}{Methods} & \multirow{2}{*}{Status} & \multicolumn{1}{c|}{Terminal} & \multicolumn{1}{c|}{HPWL} & \multicolumn{1}{c|}{GRT\_WL} & \multicolumn{1}{c|}{DRT\_WL} & \multicolumn{1}{c|}{DRC} & \multicolumn{1}{c|}{WNS} & \multicolumn{1}{c|}{TNS} & \multicolumn{1}{c|}{Power} & \multicolumn{1}{c|}{T$_{max}$} \\ 
 &  &  & (\#) & (m) & (m) & (m) & (\#) & (ns) & (ns) & (W) & (°C) \\ \hline

\multirow{2}{*}{\texttt{ariane133}} 
 & \textit{Default} & Pass & \B 4934 & \B 2.56 & \B 4.90 & \B 4.02 & 32995 & \B 0.58 & \B 0 & \B 0.237 & \B 55.48 \\
 & \textit{Padded}    & Pass & 5466 & 3.14 & 5.79 & 4.78 & \B 3331 & 0.44 & \B 0 & 0.378 & 60.54 \\ \hline

\multirow{2}{*}{\texttt{ariane136}} 
 & \textit{Default} & Pass & \B 4793 & \B 2.68 & \B 5.06 & \B 4.15 & 33836 & \B -0.93 & \B -1796 & \B 0.281 & \B 58.38 \\
 & \textit{Padded}    & Pass & 5132 & 3.25 & 5.93 & 4.89 & \B 2337 & -0.97 & -1898 & 0.327 & 59.21 \\ \hline

\multirow{2}{*}{\texttt{black\_parrot}} 
 & \textit{Default} & Pass & \B 7667 & \B 3.46 & \B 6.69 & \B 5.64 & 55884 & \B -3.63 & \B -4 & 0.432 & 64.93 \\
 & \textit{Padded}    & Pass & 8793 & 4.60 & 8.51 & 7.11 & \B 2818 & -5.27 & -19 & \B 0.399 & \B 63.24 \\ \hline

\multirow{2}{*}{\texttt{bp\_be}} 
 & \textit{Default} & Pass & \B 2702 & \B 0.72 & \B 2.49 & \B 2.27 & 28284 & -2.83 & -727 & 0.135 & 60.73 \\
 & \textit{Padded}    & Pass & 3251 & 0.89 & 2.80 & 2.52 & \B 1695 & \B -2.72 & \B -698 & \B 0.120 & \B 56.88 \\ \hline

\multirow{2}{*}{\texttt{bp\_fe}} 
 & \textit{Default} & Pass & \B 2790 & \B 0.80 & \B 2.17 & \B 2.05 & 17492 & -0.86 & -311 & \B 0.235 & 76.62 \\
 & \textit{Padded}    & Pass & 2980 & 0.91 & 2.36 & 2.21 & \B 4392 & \B -0.83 & \B -257 & 0.242 & \B 76.44 \\ \hline

\multirow{2}{*}{\texttt{bp\_multi}} 
 & \textit{Default} & Pass & \B 7151 & \B 1.68 & \B 3.59 & \B 3.02 & 8783 & \B -4.96 & -2566 & 0.702 & 84.12 \\
 & \textit{Padded}    & Pass & 7608 & 2.15 & 4.32 & 3.62 & \B 1869 & \B -4.96 & \B -2376 & \B 0.668 & \B 82.04 \\ \hline

\multirow{2}{*}{\texttt{swerv\_wrapper}} 
 & \textit{Default}    & DRT Fail & \B 5755 & \B 2.04 & \B 3.99 & - & - & - & - & - & - \\
 & \textit{Padded}    & Pass & 5871 & 2.59 & 4.69 & \B 4.07 & \B 8271 & \B -0.43 & \B -5 & \B 0.487 & \B 81.97 \\ \hline
\multicolumn{3}{|c|}{\multirow{2}{*}{Avg. Impr. \textit{Padded} vs. \textit{Default}}} & \multirow{2}{*}{-9.73\%$\textcolor{blue}{\uparrow}$} & \multirow{2}{*}{-24.17\%$\textcolor{blue}{\uparrow}$} & \multirow{2}{*}{-17.38\%$\textcolor{blue}{\uparrow}$} & \multirow{2}{*}{-16.91\%$\textcolor{blue}{\uparrow}$} & \multirow{2}{*}{87.60\%$\textcolor{red}{\downarrow}$} & \multirow{2}{*}{-11.04\%$\textcolor{blue}{\uparrow}$} & \multirow{2}{*}{2.89\%$\textcolor{red}{\downarrow}$} & \multirow{2}{*}{-9.21\%$\textcolor{blue}{\uparrow}$} & \multirow{2}{*}{0.18\%$\textcolor{red}{\downarrow}$}\\
\multicolumn{3}{|c|}{\multirow{2}{*}{}}  &  &  &  & &  &  &  &  &\\ \hline
\end{tabular}
\end{table*}

Regarding geometric and electrical metrics, the contest winners show high sensitivity to specific benchmark configurations, making it difficult to identify a universal winner based solely on HPWL or terminal counts. However, \textit{TCAD'25} consistently delivers an order-of-magnitude reduction in runtime, underscoring the efficiency of its GPU-accelerated architecture. The visual topologies in Fig.~\ref{fig:lol} highlight the diverse inter-tier integration strategies of the evaluated placers. While \textit{TCAD'25} achieves a balanced and spatially uniform cell distribution across the die pair, \textit{cadb1049} exhibits excessive local density due to its highly clustered cell patterns, which complicates routing. Notably, \textit{cadb1038} consistently allocates nearly all standard cells to the top die across all test cases, leaving the bottom die nearly vacant of logic—an extremely unbalanced partitioning strategy that underutilizes 3D stacking resources. In contrast, the balanced distribution maintained by \textit{TCAD'25} effectively alleviates routing congestion and facilitates subsequent design closure.

Importantly, our analysis reveals a divergence between geometric wirelength and final timing performance. For the \texttt{bp\_be} case, although \textit{cadb0013} and \textit{TCAD'25} yield nearly identical HPWL, their post-route timing variance exceeds $10\times$. This discrepancy suggests that pure wirelength minimization is insufficient for 3D ICs, highlighting the necessity of PPA-driven placement objectives that account for the complex electrical interactions inherent in multi-die integration.

\subsubsection{Congestion Mitigation via Cell Padding}

Routing failures in the preceding benchmarks primarily stem from localized congestion, as contest-winning placers often prioritize HPWL minimization over congestion-aware heuristics. To address this without altering placer binaries, we apply a cell padding strategy by padding standard cell widths in the \textit{lef} library by $5\times$ the site width ($\sim$0.95\,$\mu$m), explicitly reserving routing resources through a sparser distribution. 

Table~\ref{table:cell_padding} summarizes the impact of cell padding using \textit{TCAD'25} as the baseline method. As expected, padding induces a trade-off: average HPWL and routed wirelength (DRT\_WL) increase by 24.17\% and 16.91\%, respectively. However, this overhead is justified by the improvement in flow success rate, enabling \texttt{swerv\_wrapper} to achieve full-flow completion. Crucially, padding drastically improves routability, reducing DRC violations by 87.60\% on average. Regarding timing and power, the performance is mixed, with the padding strategy excelling in roughly half the cases. This split indicates that congestion, timing, and power are tightly coupled metrics; while padding reduces congestion-related detours, the resulting wirelength expansion can offset gains. Thus, these metrics must be balanced holistically to achieve optimal design quality.

\subsubsection{Evaluation of HBT Legalization under Grid Constraints}

\begin{table*}[ht!]
\renewrobustcmd{\bfseries}{\fontseries{b}\selectfont}
\renewrobustcmd{\boldmath}{}
\newrobustcmd{\B}{\bfseries}
\caption{Quantitative assessment of HBT legalization strategies under varying manufacturing pitch constraints ($2, 4, 6~\mu\text{m}$). We benchmark three assignment methodologies: greedy assignment (\textit{Greedy}), sliding window heuristic (\textit{SW}), and our implemented RL-based method (\textit{RL}). Best valid results are highlighted in \textbf{bold}.}
\label{table:hbt}
\centering
\renewcommand{\arraystretch}{1.07}
\begin{tabular}{|c|c|c|c|c|c|c|c|c|c|c|c|}
\hline
\multirow{2}{*}{Designs} & \multirow{2}{*}{Methods} & \multicolumn{1}{c|}{Pitch} & \multicolumn{1}{c|}{HBT Util.} & \multicolumn{1}{c|}{HPWL*} & \multicolumn{1}{c|}{Disp.*} & \multicolumn{1}{c|}{Runtime*} & \multicolumn{1}{c|}{GRT\_WL} & \multicolumn{1}{c|}{DRT\_WL} & \multicolumn{1}{c|}{DRC} & \multicolumn{1}{c|}{WNS} & \multicolumn{1}{c|}{TNS} \\ 
 & & ($\mu$m) & (\%) & (m) & ($10^{-3}$m) & (s) & (m) & (m) & (\#) & (ns) & (ns) \\ \hline

\multirow{5}{*}{\texttt{ariane133}} 
 &\textit{Greedy} & 2 & 2.24\% & 314.23 & 0.55 & 40 & 5.80 & 4.78 & 3768 & 0.44 & 0 \\ 
 &\textit{SW}& 4 & 9.00\% & 316.41 & 2.66 & 2982 & 5.82 & 4.80 & 1236 & 0.44 & 0 \\ 
 &\textit{RL}      & 4 & 9.00\% & \B 315.93 & \B 2.18 & \B 87 & 5.82 & 4.79 & 1246 & 0.43 & 0 \\
 &\textit{SW}& 6 & 20.32\% & 321.03 & 6.08 & 763 & 5.88 & 4.83 & 1129 & 0.44 & 0 \\ 
 &\textit{RL}      & 6 & 20.32\% & \B 319.93 & \B 5.06 & \B 89 & 5.87 & 4.82 & 1216 & 0.43 & 0 \\ \hline

\multirow{5}{*}{\texttt{ariane136}} 
 &\textit{Greedy} & 2 & 2.11\% & 325.04 & 0.51 & 38 & 5.93 & 4.89 & 2329 & -0.97 & -1880 \\ 
 &\textit{SW}& 4 & 8.45\% & 326.60 & 2.12 & 2944 & 5.95 & 4.90 & 1187 & -0.96 & -1863 \\ 
 &\textit{RL}      & 4 & 8.45\% & \B 326.02 & \B 1.55 & \B 94 & 5.95 & 4.90 & 1119 & -0.96 & -1861 \\
 &\textit{SW}& 6 & 19.08\% & 329.82 & 4.63 & 889 & 5.99 & 4.93 & 1154 & -0.97 & -1874 \\ 
 &\textit{RL}      & 6 & 19.08\% & \B 328.35 & \B 3.44 & \B 108 & 5.98 & 4.91 & 1105 & -0.97 & -1886 \\ \hline

\multirow{5}{*}{\texttt{black\_parrot}} 
 &\textit{Greedy} & 2 & 4.47\% & 460.30 & 0.88 & 50 & 8.51 & 7.12 & 2797 & -5.27 & -19 \\ 
 &\textit{SW}& 4 & 17.92\% & 462.27 & 3.04 & 2214 & 8.53 & 7.13 & 1845 & -5.27 & -19 \\ 
 &\textit{RL}      & 4 & 17.92\% & \B 461.68 & \B 2.47 & \B 107 & 8.53 & 7.12 & 1813 & -5.27 & -19 \\
 &\textit{SW}& 6 & 40.42\% & 466.24 & 6.62 & 527 & 8.58 & 7.16 & 1647 & -5.27 & -19 \\ 
 &\textit{RL}      & 6 & 40.42\% & \B 464.54 & \B 5.12 & \B 130 & 8.57 & 7.15 & 1605 & -5.27 & -19 \\ \hline

\multirow{5}{*}{\texttt{bp\_be}} 
 &\textit{Greedy} & 2 & 4.53\% & 88.99 & 0.33 & 13 & 2.80 & 2.52 & 1953 & -2.72 & -698 \\ 
 &\textit{SW}& 4 & 18.13\% & 90.39 & 1.65 & 399 & 2.82 & 2.53 & 209 & -2.72 & -698 \\ 
 &\textit{RL}      & 4 & 18.13\% & \B 89.95 & \B 1.28 & \B 60 & 2.81 & 2.53 & 211 & -2.72 & -698 \\
 &\textit{SW}& 6 & 40.95\% & 97.13 & 6.45 & \B 82 & 2.89 & 2.60 & 269 & -2.72 & -698 \\ 
 &\textit{RL}      & 6 & 40.95\% & \B 95.74 & \B 5.40 & 91 & 2.88 & 2.58 & 263 & -2.72 & -698 \\ \hline

\multirow{5}{*}{\texttt{bp\_fe}} 
 &\textit{Greedy} & 2 & 5.46\% & 91.40 & 0.30 & 11 & 2.36 & 2.21 & 4693 & -0.83 & -250 \\ 
 &\textit{SW}& 4 & 22.02\% & 93.87 & 2.93 & 228 & 2.40 & 2.24 & 459 & -0.83 & -210 \\ 
 &\textit{RL}      & 4 & 22.02\% & \B 93.86 & \B 2.30 & \B 70 & 2.40 & 2.24 & 486 & -0.83 & -238 \\
 &\textit{SW}& 6 & 49.27\% & \B 106.74 & 12.56 & \B 52 & 2.54 & 2.38 & 325 & -0.84 & -253 \\ 
 &\textit{RL}      & 6 & 49.27\% & 108.12 & \B 11.50 & 65 & 2.55 & 2.38 & 341 & -0.84 & -263 \\ \hline

\multirow{5}{*}{\texttt{bp\_multi}} 
 &\textit{Greedy} & 2 & 4.90\% & 214.82 & 0.77 & 38 & 4.33 & 3.62 & 1917 & -4.96 & -2379 \\ 
 &\textit{SW}& 4 & 19.60\% & 217.34 & 3.23 & 1420 & 4.36 & 3.64 & 47 & -4.96 & -2377 \\ 
 &\textit{RL}      & 4 & 19.60\% & \B 216.42 & \B 2.47 & \B 110 & 4.35 & 3.63 & 36 & -4.96 & -2377 \\
 &\textit{SW}& 6 & 44.33\% & 231.37 & 12.65 & 324 & 4.51 & 3.77 & 27 & -4.96 & -2402 \\ 
 &\textit{RL}      & 6 & 44.33\% & \B 227.66 & \B 10.20 & \B 26 & 4.47 & 3.74 & 26 & -4.96 & -2388 \\ \hline

\multirow{5}{*}{\texttt{swerv\_wrapper}} 
 &\textit{Greedy} & 2 & 4.33\% & 258.63 & 0.59 & 26 & 4.69 & 4.07 & 8253 & -0.43 & -5 \\ 
 &\textit{SW}& 4 & 17.33\% & 260.69 & 2.68 & 955 & 4.72 & 4.09 & 5054 & -0.43 & -5 \\ 
 &\textit{RL}      & 4 & 17.33\% & \B 259.87 & \B 2.03 & \B 82 & 4.70 & 4.08 & 5046 & -0.43 & -5 \\ 
 &\textit{SW}& 6 & 39.31\% & 271.76 & 9.98 & 260 & 4.84 & 4.19 & 4710 & -0.43 & -5 \\ 
 &\textit{RL}      & 6 & 39.31\% & \B 267.49 & \B 7.48 & \B 140 & 4.79 & 4.15 & 140 & -0.43 & -6 \\ \hline
 
\multicolumn{2}{|c|}{\multirow{2}{*}{Avg. Impr. \textit{RL} vs. \textit{SW}}} & \multirow{2}{*}{4} & \multirow{2}{*}{-} & \multirow{2}{*}{0.24\%$\textcolor{red}{\downarrow}$} & \multirow{2}{*}{22.06\%$\textcolor{red}{\downarrow}$} & \multirow{2}{*}{89.57\%$\textcolor{red}{\downarrow}$} & \multirow{2}{*}{-} & \multirow{2}{*}{-} & \multirow{2}{*}{-} & \multirow{2}{*}{-} & \multirow{2}{*}{-}\\
\multicolumn{2}{|c|}{\multirow{2}{*}{}} &  &  &  &  & &  &  & & &\\ \hline

\multicolumn{2}{|c|}{\multirow{2}{*}{Avg. Impr. \textit{RL} vs. \textit{SW}}} & \multirow{2}{*}{6} & \multirow{2}{*}{-} & \multirow{2}{*}{0.64\%$\textcolor{red}{\downarrow}$} & \multirow{2}{*}{19.23\%$\textcolor{red}{\downarrow}$} & \multirow{2}{*}{45.54\%$\textcolor{red}{\downarrow}$} & \multirow{2}{*}{-} & \multirow{2}{*}{-} & \multirow{2}{*}{-} & \multirow{2}{*}{-} & \multirow{2}{*}{-}\\
\multicolumn{2}{|c|}{\multirow{2}{*}{}} &  &  &  &  & &  &  & & &\\ \hline
\end{tabular}
\begin{tablenotes}
\item *Average improvement is calculated exclusively for HPWL, Disp., and Runtime, as these represent the primary metrics for assessing the physical quality and computational efficiency of the legalization methodologies. Other metrics, such as routed wirelength and timing slacks, are significantly influenced by downstream routing engines.
\end{tablenotes}
\vspace{-2em}
\end{table*}

\begin{figure}[ht!]
\centering
\includegraphics[width=\linewidth, keepaspectratio]{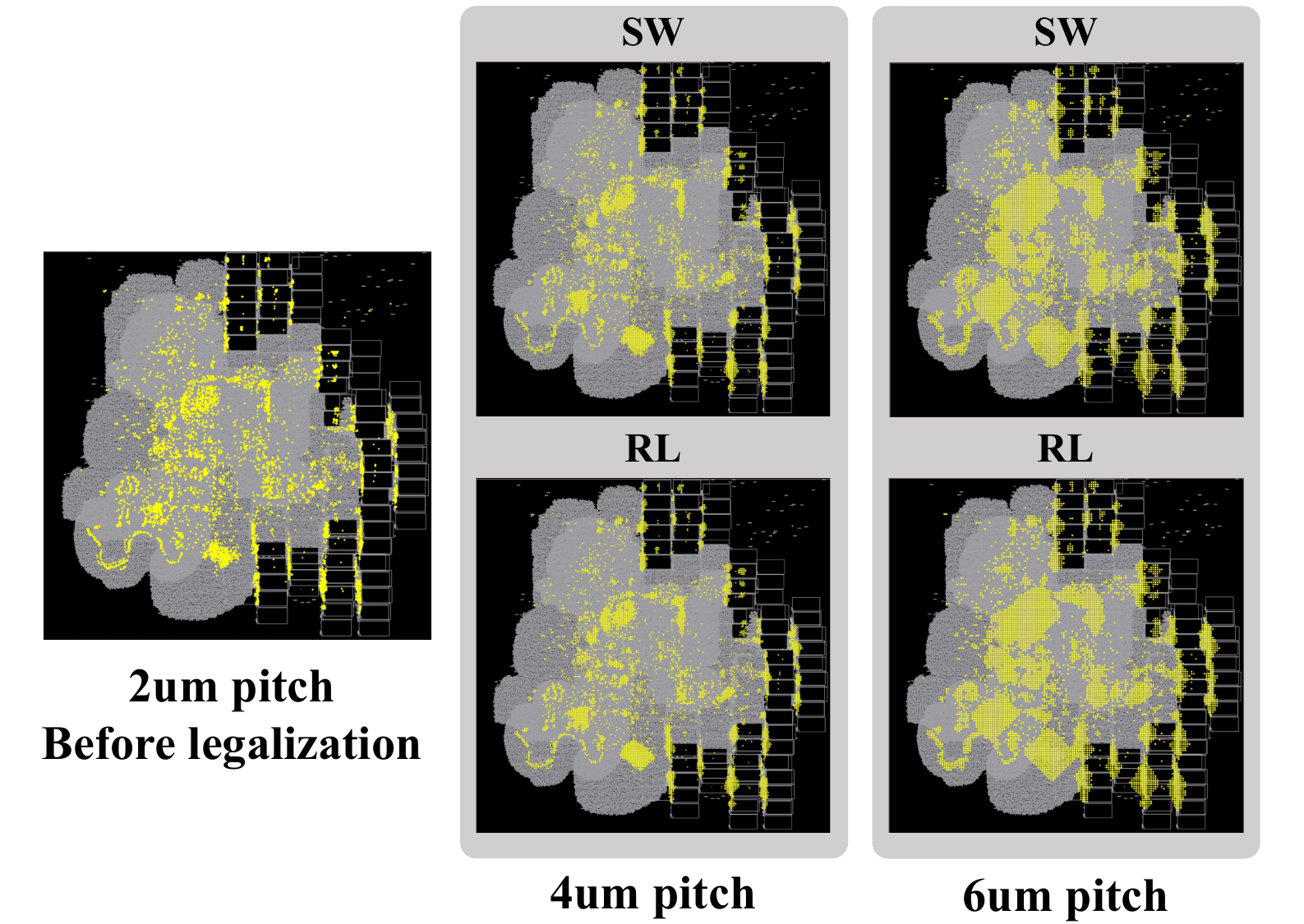}
\caption{Visualization of HBT distribution for the \texttt{ariane133} benchmark. The yellow markers illustrate the HBT assigned locations from initial placement to legalized grids.}
\label{fig:hbt}
\vspace{-2em}
\end{figure}

While initial placement coordinates satisfy basic spacing constraints, practical manufacturing requires HBTs to be aligned with coarse, predefined grids. To evaluate legalization robustness under these stricter conditions, we snap initial HBT positions to manufacturing grids with pitches of $2\,\mu$m, $4\,\mu$m, and $6\,\mu$m. We compare a baseline \textit{Greedy} strategy (sufficient for the $2\,\mu$m pitch) against a sliding window heuristic (\textit{SW})~\cite{pentapati2023vialegal} and our reinforcement learning-based approach (\textit{RL}) for coarser pitches where resource contention occurs. Quantitative results and layout visualizations are provided in Table~\ref{table:hbt} and Fig.~\ref{fig:hbt}, respectively.

As shown in Table~\ref{table:hbt}, \textit{RL} consistently maintains superior solution quality over \textit{SW}, reducing average HBT displacement by 22.06\% and 19.23\% under $4\,\mu$m and $6\,\mu$m constraints, respectively. This improved preservation of placement intent directly correlates with lower HPWL. Analysis of PPA trends reveals that timing performance (WNS/TNS) remains largely orthogonal to pitch constraints. In contrast, routability improves significantly at coarser pitches; the reduced HBT density at $6\,\mu$m alleviates local congestion, leading to a sharp decline in DRC violations. Notably, for \texttt{swerv\_wrapper} at $6\,\mu$m, \textit{RL} effectively resolves congestion bottlenecks, reducing DRC violations from 4710 to 140 compared to \textit{SW}.

The most significant advantage of the \textit{RL} method lies in its computational efficiency and scalability. For the \texttt{ariane133} benchmark under a $4\,\mu$m constraint, \textit{RL} achieves an 18.23\% displacement reduction while operating $34\times$ faster than \textit{SW} (87s vs. 2982s). Even under the $6\,\mu$m constraint with a smaller search space, \textit{RL} maintains an $8\times$ speedup (89s vs. 763s). This performance gain stems from the algorithmic contrast between the two: while \textit{SW} exhaustively traverses the entire canvas, \textit{RL} intelligently focuses on regional refinement, minimizing overhead while maximizing placement quality.

\subsubsection{Benchmarking with ICCAD 2022 Macro-Free Contest Winners}

\begin{table*}[t]
\renewrobustcmd{\bfseries}{\fontseries{b}\selectfont}
\renewrobustcmd{\boldmath}{}
\newrobustcmd{\B}{\bfseries}
\caption{Quantitative backend performance evaluation for LoL integration using macro-free benchmarks adapted for the ICCAD 2022 contest protocol. We evaluate binaries from the contest winners: 1st place \textit{cadb1051}~\cite{zhao2023ipl3d} and 3rd place \textit{cadb1021}~\cite{liao2024bistratal}.}
\label{table:iccad2022}
\centering
\renewcommand{\arraystretch}{1.12}
\begin{tabular}{|c|c|c|c|c|c|c|c|c|c|c|c|}
\hline
\multirow{2}{*}{Designs} & \multirow{2}{*}{Methods} & \multirow{2}{*}{Status} & \multicolumn{1}{c|}{Terminal} & \multicolumn{1}{c|}{HPWL} & \multicolumn{1}{c|}{Runtime} & \multicolumn{1}{c|}{GRT\_WL} & \multicolumn{1}{c|}{DRT\_WL} & \multicolumn{1}{c|}{WNS} & \multicolumn{1}{c|}{TNS} & \multicolumn{1}{c|}{Power} & \multicolumn{1}{c|}{T$_{max}$} \\ 
 &  &  & (\#) & (m) & (s) & (m) & (m) & (ns) & (ns) & (W) & (°C) \\ \hline

\multirow{2}{*}{\texttt{aes}} 
 & \textit{cadb1051} & Pass & 2883 & 0.110 & 54 & 0.270 & 0.215 & -0.177 & -8.46 & 0.411 & 647.66 \\ 
 & \textit{cadb1021} & Pass & 1492 & 0.115 & 36 & 0.276 & 0.208 & -0.238 & -14.36 & 0.389 & 613.89 \\ \hline

\multirow{2}{*}{\texttt{dynamic\_node}} 
 & \textit{cadb1051} & Pass & 2716 & 0.074 & 51 & 0.260 & 0.233 & -0.053 & -0.05 & 0.012 & 60.17 \\ 
 & \textit{cadb1021} & Pass & 3083 & 0.087 & 37 & 0.287 & 0.258 & 0.228 & 0.0 & 0.011 & 58.96 \\ \hline

\multirow{2}{*}{\texttt{ibex}} 
 & \textit{cadb1051} & PL Fail & - & - & - & - & - & - & - & - & - \\ 
 & \textit{cadb1021} & Pass & 1600 & 0.164 & 38 & 0.354 & 0.271 & -0.085 & -9.27 & 0.096 & 101.72 \\ \hline

\multirow{2}{*}{\texttt{jpeg}} 
 & \textit{cadb1051} & PL Fail & - & - & - & - & - & - & - & - & - \\ 
 & \textit{cadb1021} & Pass & 13458 & 0.420 & 229 & 0.792 & 0.586 & -0.273 & -53.11 & 0.166 & 94.17 \\ \hline

\multirow{2}{*}{\texttt{swerv}} 
 & \textit{cadb1051} & PL Fail & - & - & - & - & - & - & - & - & - \\ 
 & \textit{cadb1021} & Pass & 11285 & 1.260 & 161 & 2.714 & 2.300 & -0.918 & -368.90 & 0.220 & 89.93 \\ \hline

\end{tabular}
\end{table*}

To further validate the framework, we evaluate macro-free LoL scenarios following the ICCAD 2022 protocol using five designs. As summarized in Table~\ref{table:iccad2022}, \textit{cadb1021}~\cite{liao2024bistratal} demonstrates superior robustness by successfully passing all test cases, whereas \textit{cadb1051}~\cite{zhao2023ipl3d} encounters three placement failures. Notably, the thermal anomaly observed in the \texttt{aes} benchmark is identified as a characteristic of its high power density rather than an artifact of our 3D framework, as the power metrics remain consistent with the 2D OpenROAD baseline.

\section{Conclusion and Future Work}

This work introduces Open3DBench, a comprehensive open-source 3D-IC backend benchmark suite. The methodology introduces a memory-on-logic flow that incorporates GNN-based partitioning and specialized macro placement strategies considering regularity and wirelength. We also provide a logic-on-logic framework that enables the standardized evaluation of ICCAD contest binaries within a complete backend implementation. Furthermore, the work integrates an RL-based HBT legalization algorithm to resolve constraints for industrial manufacturability. By offering PDK compatibility and standardized workflows, we hope Open3DBench serves as a foundational catalyst for the 3D-IC research community, fostering the development and rigorous validation of next-generation physical design algorithms.

In the future, we plan to extend Open3DBench to support heterogeneous integration and more advanced technology nodes. We also intend to incorporate comprehensive comparisons between our open-source flows and commercial EDA tool-based results to further enhance the empirical persuasiveness of the suite.

\section{Acknowledgment}

The authors would like to declare the use of Google Gemini~3.0~\footnote{https://gemini.google.com/app} to improve the initial text of this manuscript. The tool was exclusively used to enhance the readability, grammatical accuracy, and logical coherence of the article.

\bibliographystyle{IEEEtran}
\bibliography{main}

@string{DAC = "Proc. DAC"}

@string{ASPDAC = "Proc. ASP-DAC"}

@string{ICCAD = "Proc. ICCAD"}

@string{TCAD = "IEEE Trans. CAD"}

@string{ISPD = "Proc. ISPD"}

@string{GOMATech = "Proc. GOMATech"}

@string{TVLSI = "IEEE Trans. VLSI"}

@string{DATE = "Proc. DATE"}

@string{MLCAD = "Proc. MLCAD"}

@string{ISLPED = "Proc. ISLPED"}

@string{ISSCC = "Proc. ISSCC"}

@string{IEDM = "Proc. IEDM"}

@IEEEtranBSTCTL{IEEEexample:BSTcontrol,
  CTLuse_forced_etal       = "yes",
  CTLmax_names_forced_etal = "3",
  CTLnames_show_etal       = "1",
  CTLdash_repeated_names = "no"
}

@inproceedings{lin2019dreamplace,
  title={{DREAMPlace}: Deep learning toolkit-enabled {GPU} acceleration for modern {VLSI} placement},
  author={Lin, Yibo and Dhar, Shounak and Li, Wuxi and Ren, Haoxing and Khailany, Brucek and Pan, David Z},
  booktitle=DAC,
  pages={1--6},
  year={2019}
}

@inproceedings{zhao2025physical,
  title={Physical Design for Advanced 3{D} {IC}s: Challenges and Solutions},
  author={Zhao, Yunxuan and Zou, Lancheng and Yu, Bei},
  booktitle=ISPD,
  year={2025},
  pages={1--9}
}

@ARTICLE{Moore1998cramming,
  author={Moore, G.E.},
  journal={Proc. IEEE}, 
  title={Cramming More Components Onto Integrated Circuits}, 
  year={1998},
  volume={86},
  number={1},
  pages={82--85}
}

@inproceedings{hu2023benchmark,
  author={Hu, Kai-Shun and Chi, Hao-Yu and Lin, I-Jye and Wu, Yi-Hsuan and Chen, Wei-Hsu and Hsieh, Yi-Ting},
  booktitle=ICCAD, 
  title={2023 {ICCAD} {CAD} Contest Problem {B}: 3{D} Placement with Macros}, 
  year={2023},
  pages={1--6}
}

@inproceedings{hu2022benchmark,
author = {Hu, Kai-Shun and Lin, I-Jye and Huang, Yu-Hui and Chi, Hao-Yu and Wu, Yi-Hsuan and Shen, Chin-Fang Cindy},
title = {2022 {ICCAD} {CAD} Contest Problem {B}: 3{D} Placement with {D2D} Vertical Connections},
year = {2022},
pages = {1--5},
booktitle = ICCAD
}

@inproceedings{chang2016cascade2d,
  title={Cascade2{D}: A design-aware partitioning approach to monolithic {3D IC} with {2D} commercial tools},
  author={Chang, Kyungwook and Sinha, Saurabh and Cline, Brian and Southerland, Raney and Doherty, Michael and Yeric, Greg and Lim, Sung Kyu},
  booktitle=ICCAD,
  pages={1--8},
  year={2016}
}

@article{panth2017shrunk,
  title={{Shrunk-2-D}: A physical design methodology to build commercial-quality monolithic {3-D ICs}},
  author={Panth, Shreepad and Samadi, Kambiz and Du, Yang and Lim, Sung Kyu},
  journal=TCAD,
  volume={36},
  number={10},
  pages={1716--1724},
  year={2017}
}

@INPROCEEDINGS{panth2014shrunk,
  author={Panth, Shreepad and Samadi, Kambiz and Du, Yang and Lim, Sung Kyu},
  booktitle=ISLPED, 
  title={Design and CAD methodologies for low power gate-level monolithic 3D ICs}, 
  year={2014},
  volume={},
  number={},
  pages={171--176},
}

@inproceedings{ku2018compact,
  title={Compact-{2D}: A physical design methodology to build commercial-quality face-to-face-bonded {3D ICs}},
  author={Ku, Bon Woong and Chang, Kyungwook and Lim, Sung Kyu},
  booktitle=ISPD,
  pages={90--97},
  year={2018}
}

@inproceedings{bamberg2020macro,
  title={Macro-3{D}: A physical design methodology for face-to-face-stacked heterogeneous {3D} {ICs}},
  author={Bamberg, Lennart and Garc{\'\i}a-Ortiz, Alberto and Zhu, Lingjun and Pentapati, Sai and Lim, Sung Kyu and others},
  booktitle=DATE,
  pages={37--42},
  year={2020}
}

@inproceedings{pentapati2020pin,
  title={{Pin-3D}: A physical synthesis and post-layout optimization flow for heterogeneous monolithic {3D ICs}},
  author={Pentapati, Sai Surya Kiran and Chang, Kyungwook and Gerousis, Vassilios and Sengupta, Rwik and Lim, Sung Kyu},
  booktitle=ICCAD,
  pages={1--9},
  year={2020}
}

@inproceedings{vanna2021snap,
  title={{Snap-3D}: A constrained placement-driven physical design methodology for face-to-face-bonded {3D ICs}},
  author={Vanna-Iampikul, Pruek and Shao, Chengjia and Lu, Yi-Chen and Pentapati, Sai and Lim, Sung Kyu},
  booktitle=ISPD,
  pages={39--46},
  year={2021}
}

@inproceedings{ajayi2019openroad,
  title={Open{ROAD}: Toward a self-driving, open-source digital layout implementation tool chain},
  author={Ajayi, T and Blaauw, D and Chan, TB and Cheng, CK and Chhabria, VA and Choo, DK and Coltella, M and Dobre, S and Dreslinski, R and Foga{\c{c}}a, M and others},
  booktitle=GOMATech,
  year={2019},
  pages={1105--1110}
}

@inproceedings{pu2024incremacro,
  title={{IncreMacro}: Incremental Macro Placement Refinement},
  author={Pu, Yuan and Chen, Tinghuan and He, Zhuolun and Bai, Chen and Zheng, Haisheng and Lin, Yibo and Yu, Bei},
  booktitle=ISPD,
  pages={169--176},
  year={2024}
}

@INPROCEEDINGS{chen2023stronger,
  author={Chen, Yifan and Wen, Zaiwen and Liang, Yun and Lin, Yibo},
  booktitle=ICCAD, 
  title={Stronger Mixed-Size Placement Backbone Considering Second-Order Information}, 
  year={2023},
  pages={1--9}
}

@article{liao2024analytical,
author = {Liao, Peiyu and Zhao, Yuxuan and Guo, Dawei and Lin, Yibo and Yu, Bei},
title = {Analytical Die-to-Die {3-D} Placement With Bistratal Wirelength Model and {GPU} Acceleration},
year = {2024},
volume = {43},
number = {6},
journal = TCAD,
pages = {1624--1637}
}

@article{hsu2013tsv,
  title={{TSV}-aware analytical placement for {3-D} {IC} designs based on a novel weighted-average wirelength model},
  author={Hsu, Meng-Kai and Balabanov, Valeriy and Chang, Yao-Wen},
  journal=TCAD,
  volume={32},
  number={4},
  pages={497--509},
  year={2013}
}

@inproceedings{kim2025ta3d,
author = {Kim, Donggyu and Kim, Minjae and Hur, Junseok and Lee, Jakang and Cho, Jinoh and Kang, Seokhyeong},
title = {{TA3D}: Timing-Aware {3D IC} Partitioning and Placement by Optimizing the Critical Path},
year = {2024},
booktitle = MLCAD,
pages={1--7}
}

@misc{wolf2013yosys,
  author       = {Wolf, Clifford and Glaser, Johann and Kepler, Johannes},
  title        = {Yosys},
  howpublished = {\url{https://github.com/YosysHQ/yosys}}
}

@misc{nangate45,
  author       = {Rob, Aslett},
  title        = {Nangate Technology Node},
  howpublished = {\url{https://si2.org/open-cell-library/}}
}

@article{wei2011skyline,
  title={A skyline heuristic for the {2D} rectangular packing and strip packing problems},
  author={Wei, Lijun and Oon, Wee-Chong and Zhu, Wenbin and Lim, Andrew},
  journal={European Journal of Operational Research},
  volume={215},
  number={2},
  pages={337--346},
  year={2011},
}

@inproceedings{liu2024hbt,
author = {Liu, Siting and Jiang, Jiaxi and He, Zhuolun and Wang, Ziyi and Lin, Yibo and Yu, Bei and Wong, Martin},
title = {Routing-aware Legal Hybrid Bonding Terminal Assignment for {3D} Face-to-Face Stacked {ICs}},
year = {2024},
booktitle = ISPD,
pages = {75--82}
}

@inproceedings{pan2006fast,
  author={Pan, Min and Chu, Chris},
  booktitle=ICCAD, 
  title={{FastRoute}: A Step to Integrate Global Routing into Placement}, 
  year={2006},
  pages={464--471}
}

@ARTICLE{Kahng2021triton,
  author={Kahng, Andrew B. and Wang, Lutong and Xu, Bangqi},
  journal=TCAD, 
  title={{TritonRoute}: The Open-Source Detailed Router}, 
  year={2021},
  volume={40},
  number={3},
  pages={547--559}
}

@misc{openrcx,
  author       = {},
  title        = {Open{RCX}},
  howpublished = {\url{https://github.com/The-OpenROAD-Project/OpenRCX}}
}

@INPROCEEDINGS{han2022hotspot,
  author={Han, Jun-Han and Guo, Xinfei and Skadron, Kevin and Stan, Mircea R.},
  booktitle={Proc. IEEE Intersociety Conference on Thermal and Thermomechanical Phenomena in Electronic Systems}, 
  title={From 2.5{D} to 3{D} Chiplet Systems: Investigation of Thermal Implications with {HotSpot} 7.0}, 
  year={2022},
  pages={1--6}
}

@misc{hotspot,
  author       = {},
  title        = {Hot{S}pot},
  howpublished = {\url{https://github.com/uvahotspot/HotSpot}}
}

@ARTICLE{kahng2024hier,
  author={Kahng, Andrew B. and Varadarajan, Ravi and Wang, Zhiang},
  journal=TCAD, 
  title={Hier-{RTLMP}: A Hierarchical Automatic Macro Placer for Large-Scale Complex {IP} Blocks}, 
  year={2024},
  volume={43},
  number={5},
  pages={1552--1565}
}

@inproceedings{murali2022art,
    author = {Murali, Gauthaman and Shaji, Sandra Maria and Agnesina, Anthony and Luo, Guojie and Lim, Sung Kyu},
    title = {ART-{3D}: Analytical {3D} Placement with Reinforced Parameter Tuning for Monolithic {3D} {ICs}},
    year = {2022},
    booktitle = ISPD,
    pages = {97--104}
}

@INPROCEEDINGS{meta2024isscc,
  author={Wu, Tony F. and Liu, Huichu and Sumbul, H. Ekin and Yang, Lita and Baheti, Dipti and Coriell, Jeremy and Koven, William and Krishnan, Anu and Mittal, Mohit and Moreira, Matheus Trevisan and Waugaman, Max and Ye, Laurent and Beigné, Edith},
  booktitle=ISSCC,
  title={A {3D} integrated Prototype System-on-Chip for Augmented Reality Applications Using Face-to-Face Wafer Bonded 7nm Logic at 2$\mu$m Pitch with up to 40\% Energy Reduction at Iso-Area Footprint}, 
  year={2024},
  volume={67},
  number={},
  pages={210--212}
}

@misc{integrity3dic,
  author = {{Cadence Design Systems}},
  title = {{Integrity 3D-IC Platform}},
  year = {2024},
  howpublished = {\url{https://www.cadence.com/en_US/home/tools/digital-design-and-signoff/soc-implementation-and-floorplanning/integrity-3dic-platform.html}}
}

@misc{3diccompiler,
  author = {{Synopsys}},
  title = {{3DIC Compiler: Platform for Multi-Die Designs}},
  year = {2025},
  howpublished = {\url{https://www.synopsys.com/implementation-and-signoff/3dic-design.html}}
}

@INPROCEEDINGS{lu2020tpgnn,
  author={Lu, Yi-Chen and Kiran Pentapati, Sai Surya and Zhu, Lingjun and Samadi, Kambiz and Lim, Sung Kyu},
  booktitle=DAC, 
  title={{TP-GNN}: A Graph Neural Network Framework for Tier Partitioning in Monolithic {3D ICs}}, 
  year={2020},
  volume={},
  number={},
  pages={1--6},
}

@inproceedings{jeong2025ppa,
author = {Jeong, Eunsol and Kim, Taewhan and Park, Heechun},
title = {{PPA}-Aware Tier Partitioning for {3D IC} Placement with {ILP} Formulation},
year = {2025},
booktitle = ASPDAC,
pages = {879--885},
}

@inproceedings{huang2025snake,
author = {Huang, Yen-Hsiang and Lim, Sung},
year = {2025},
pages = {1--9},
title = {{Snake-3D}: Differentiable Learning for Cross-Tier Logic Path Snaking Optimization in {3D ICs}},
booktitle = ICCAD
}

@inproceedings{ku2020pin-in-the-middile,
author = {Ku, Bon Woong and Lim, Sung Kyu},
title = {Pin-in-the-middle: An efficient block pin assignment methodology for block-level monolithic {3D IC}s},
year = {2020},
booktitle = ISLPED,
pages = {85--90}
}

@inproceedings{agnesina2022hier3d,
author = {Agnesina, Anthony and Brunion, Moritz and Garcia-Ortiz, Alberto and Catthoor, Francky and Milojevic, Dragomir and Komalan, Manu and Cavalcante, Matheus and Riedel, Samuel and Benini, Luca and Lim, Sung Kyu},
title = {Hier-3D: A Hierarchical Physical Design Methodology for Face-to-Face-Bonded 3D ICs},
year = {2022},
booktitle = ISLPED,
pages = {1--6}
}

@inproceedings{park2025closing,
author = {Park, Min and Vanna-iampikul, Pruek and Lim, Sung},
year = {2025},
pages = {1-9},
title = {Closing the Gap: Advantages of Block-Level over Gate-Level in {3D IC} Design for Advanced Nodes},
booktitle = ICCAD
}

@INPROCEEDINGS{Hsiao2025dco3d,
  author={Hsiao, Hao-Hsiang and Lu, Yi-Chen and Vanna-Iampikul, Pruek and Agnesina, Anthony and Liang, Rongjian and Lu, Yuan-Hsiang and Ren, Haoxing and Lim, Sung Kyu},
  booktitle=DAC, 
  title={{DCO-3D}: Differentiable Congestion Optimization in {3D ICs}}, 
  year={2025},
  volume={},
  number={},
  pages={1--7},
}

@ARTICLE{pentapati2022metal,
  author={Pentapati, Sai and Lim, Sung Kyu},
  journal=TVLSI, 
  title={Metal Layer Sharing: A Routing Optimization Technique for Monolithic {3D ICs}}, 
  year={2022},
  volume={30},
  number={9},
  pages={1355--1367},
}

@INPROCEEDINGS{hu2025gnnmls,
  author={Hu, Jiawei and Vanna-Iampikul, Pruek and Zhuang, Zhen and Ho, Tsung-Yi and Lim, Sung Kyu},
  booktitle=DAC, 
  title={{GNN-MLS}: Signal Routing in Mixed-Node {3D IC}s through {GNN}-Assisted Metal Layer Sharing}, 
  year={2025},
  volume={},
  number={},
  pages={1--7},
}

@inproceedings{vanna2025net-to-pad,
author = {Vanna-iampikul, Pruek and Yoon, Junsik and Park, Chaeryung and Yeap, Gary and Lim, Sung Kyu},
title = {Placement-Aware {3D} Net-to-Pad Assignment for Array-Style Hybrid Bonding {3D ICs}},
year = {2025},
booktitle = ISPD,
pages = {200--208},
}

@inproceedings{pentapati2023vialegal,
author = {Pentapati, Sai and Agnesina, Anthony and Brunion, Moritz and Huang, Yen-Hsiang and Lim, Sung Kyu},
title = {On Legalization of Die Bonding Bumps and Pads for {3D IC}s},
year = {2023},
booktitle = ISPD,
pages = {62--70},
}

@inproceedings{kim2025unified,
author = {Kim, Gyumin and Park, Heechun},
year = {2025},
pages = {1--9},
title = {A Unified Design Flow for Homogeneous and Heterogeneous {3D} Integration with Fine-Pitch Hybrid Bonding},
booktitle = ICCAD
}

@INPROCEEDINGS{cong2007t3place,
  author={Cong, Jason and Luo, Guojie and Wei, Jie and Zhang, Yan},
  booktitle=ASPDAC, 
  title={Thermal-Aware {3D IC} Placement Via Transformation}, 
  year={2007},
  volume={},
  number={},
  pages={780--785},
}

@inproceedings{cong2010analytical,
author = {Cong, Jason and Luo, Guojie},
title = {An analytical placer for mixed-size {3D} placement},
year = {2010},
booktitle = ISPD,
pages = {61--66},
}

@ARTICLE{hsu2013ntuplace3-3d,
  author={Hsu, Meng-Kai and Balabanov, Valeriy and Chang, Yao-Wen},
  journal=TCAD, 
  title={{TSV}-Aware Analytical Placement for {3-D IC} Designs Based on a Novel Weighted-Average Wirelength Model}, 
  year={2013},
  volume={32},
  number={4},
  pages={497--509},
}

@inproceedings{lu2016eplace3d,
author = {Lu, Jingwei and Zhuang, Hao and Kang, Ilgweon and Chen, Pengwen and Cheng, Chung-Kuan},
title = {ePlace-3D: Electrostatics based Placement for 3D-ICs},
year = {2016},
booktitle = ISPD,
pages = {11--18},
}

@INPROCEEDINGS{lin2023hyplace-3d,
  author={Lin, Jai-Ming and Lin, Yu-Chien and Kung, Hsuan and Lin, Wei-Yuan},
  booktitle=ICCAD, 
  title={HyPlace-{3D}: A Hybrid Placement Approach for {3D ICs} Using Space Transformation Technique}, 
  year={2023},
  volume={},
  number={},
  pages={1--8},
}

@ARTICLE{liao2024bistratal,
  author={Liao, Peiyu and Zhao, Yuxuan and Guo, Dawei and Lin, Yibo and Yu, Bei},
  journal=TCAD, 
  title={Analytical Die-to-Die {3-D} Placement With Bistratal Wirelength Model and {GPU} Acceleration}, 
  year={2024},
  volume={43},
  number={6},
  pages={1624--1637},
}

@INPROCEEDINGS{zhao2023ipl3d,
  author={Zhao, Xueyan and Chen, Shijian and Qiu, Yihang and Li, Jiangkao and Huang, Zhipeng and Xie, Biwei and Li, Xingquan and Bao, Yungang},
  booktitle=ICCAD, 
  title={{iPL-3D}: A Novel Bilevel Programming Model for Die-to-Die Placement}, 
  year={2023},
  volume={},
  number={},
  pages={1--9},
}

@inproceedings{chen2024mixed,
author = {Chen, Yan-Jen and Hsieh, Cheng-Hsiu and Su, Po-Han and Chen, Shao-Hsiang and Chang, Yao-Wen},
title = {Mixed-Size {3D} Analytical Placement with Heterogeneous Technology Nodes},
year = {2024},
booktitle = DAC,
pages = {1--6}
}

@ARTICLE{zhao2025analytical,
  author={Zhao, Yuxuan and Liao, Peiyu and Liu, Siting and Jiang, Jiaxi and Lin, Yibo and Yu, Bei},
  journal=TCAD, 
  title={Analytical Heterogeneous Die-to-Die {3-D} Placement With Macros}, 
  year={2025},
  volume={44},
  number={2},
  pages={402--415},
}

@inproceedings{liao2025ultrafast,
author = {Liao, Peiyu and Zhao, Yuxuan and Liu, Siting and Yu, Bei},
year = {2025},
month = {10},
pages = {1--9},
title = {Ultrafast Density Gradient Accumulation in {3D} Analytical Placement with Divergence Theorem},
booktitle = ICCAD
}

@inproceedings{zhao2025h3d,
author = {Zhao, Yuxuan and Gu, Feng and Liu, Siting and Liao, Peiyu and Yu, Bei},
year = {2025},
month = {10},
pages = {1--9},
title = {{H3D}: Heterogeneous Resources Aware Global Router for Face-to-Face Bonded {3D IC}s},
booktitle = ICCAD
}

@inproceedings{zhao20253dflow,
author = {Zhao, Yuxuan and Liao, Peiyu and Yu, Bei},
title = {{3D}-{Flow}: Flow-Based Standard Cell Legalization for {3D ICs}},
year = {2025},
booktitle = DAC,
pages = {1--7}
}

@inproceedings{chang20253ddrc,
author = {Chang, Shunjie and Wu, Youran and Chen, Jianli and Yu, Jun},
year = {2025},
month = {10},
pages = {1--9},
title = {{3D DRC}: Design Rule Checking for {3D IC} with {U-Net}-based Non-{Manhattan} Optimization},
booktitle = ICCAD
}

@ARTICLE{zhu2025open3dflow-journal,
  author={Zhu, Yifei and Luan, Zhenxuan and Feng, Dawei and Chen, Weiwei and Ren, Lei and Tan, Zhangxi},
  journal={IEEE Open Journal of Circuits and Systems}, 
  title={Revolutionize {3D}-Chip Design With {Open3DFlow}, an Open-Source {AI}-Enhanced Solution}, 
  year={2025},
  volume={6},
  number={},
  pages={169--180},
}

@inproceedings{zhu2025open3dflow-conference,
author = {Zhu, Yifei and Feng, Dawei and Luan, Zhenxuan and Ren, Lei and Chen, Weiwei and Tan, Zhangxi},
year = {2025},
month = {10},
pages = {1--9},
title = {{Open3DFlow}: An Open-Source {EDA} Platform for {3D} Chip Design with {AI} Enhancement},
booktitle = ICCAD
}

@article{martello1987hungarian,
title = {Linear Assignment Problems},
volume = {132},
pages = {259--282},
year = {1987},
journal = {Surveys in Combinatorial Optimization},
author = {Silvano Martello and Paolo Toth},
}

@ARTICLE{cheng2019replace,
  author={Cheng, Chung-Kuan and Kahng, Andrew B. and Kang, Ilgweon and Wang, Lutong},
  journal=TCAD, 
  title={{RePlAce}: Advancing Solution Quality and Routability Validation in Global Placement}, 
  year={2019},
  volume={38},
  number={9},
  pages={1717--1730},
}

@misc{opensta,
  author       = {Cherry, James and {Parallax Software, Inc.}},
  title        = {{OpenSTA}: Parallax Static Timing Analyzer},
  year         = {2023},
  howpublished = {\url{https://github.com/parallaxsw/OpenSTA}}
}

@INPROCEEDINGS{chen20203dsram,
  author={Chen, R. and Weckx, P. and Salahuddin, S. M. and Kim, S.-W. and Sisto, G. and Van der Plas, G. and Stucchi, M. and Baert, R. and Debacker, P. and Na, M.H. and Ryckaert, J. and Milojevic, D. and Beyne, E.},
  booktitle=IEDM, 
  title={{3D}-optimized {SRAM} Macro Design and Application to Memory-on-Logic {3D-IC} at Advanced Nodes}, 
  year={2020},
  volume={},
  number={},
  pages={15.2.1-15.2.4},
}

@INPROCEEDINGS{wuu20223dvcache,
  author={Wuu, John and Agarwal, Rahul and Ciraula, Michael and Dietz, Carl and Johnson, Brett and Johnson, Dave and Schreiber, Russell and Swaminathan, Raja and Walker, Will and Naffziger, Samuel},
  booktitle=ISSCC, 
  title={3{D} {V-Cache}: The Implementation of a Hybrid-Bonded {64MB} Stacked Cache for a 7nm x86-64 {CPU}}, 
  year={2022},
  volume={65},
  number={},
  pages={428--429},
}

@article{mutlu2025processingmemory,
      title={A Modern Primer on Processing in Memory}, 
      author={Onur Mutlu and Saugata Ghose and Juan Gómez-Luna and Rachata Ausavarungnirun and Mohammad Sadrosadati and Geraldo F. Oliveira},
      year={2025},
      journal={arXiv preprint: 2012.03112},
}

@inproceedings{yang20243dstacked,
author = {Yang, Lita and Kao, Changjung and Srikanth, Sriseshan and Sumbul, H. Ekin and Wu, Tony F. and Liu, Huichu and Beigne, Edith},
title = {Characterization and Design of {3D}-Stacked Memory for Image Signal Processing on {AR/VR} Devices},
year = {2024},
booktitle = {Proc. International Symposium on Memory Systems},
pages = {38--44},
}

@INPROCEEDINGS{ingerly2019foveros,
  author={Ingerly, D. B. and Amin, S. and Aryasomayajula, L. and Balankutty, A. and Borst, D. and Chandra, A. and Cheemalapati, K. and Cook, C. S. and Criss, R. and Enamul, K. and Gomes, W. and Jones, D. and Kolluru, K. C. and Kandas, A. and Kim, G.-S. and Ma, H. and Pantuso, D. and Petersburg, C.F. and Phen-givoni, M. and Pillai, A. M. and Sairam, A. and Shekhar, P. and Sinha, P. and Stover, P. and Telang, A. and Zell, Z.},
  booktitle=IEDM, 
  title={Foveros: {3D} Integration and the use of Face-to-Face Chip Stacking for Logic Devices}, 
  year={2019},
  volume={},
  number={},
  pages={19.6.1-19.6.4},
}

@INPROCEEDINGS{haruta20173layer,
  author={Haruta, Tsutomu and Nakajima, Tsutomu and Hashizume, Jun and Umebayashi, Taku and Takahashi, Hiroshi and Taniguchi, Kazuo and Kuroda, Masami and Sumihiro, Hiroshi and Enoki, Koji and Yamasaki, Takatsugu and Ikezawa, Katsuya and Kitahara, Atsushi and Zen, Masao and Oyama, Masafumi and Koga, Hiroki and Tsugawa, Hidenobu and Ogita, Tomoharu and Nagano, Takashi and Takano, Satoshi and Nomoto, Tetsuo},
  booktitle=ISSCC, 
  title={A 1/2.3inch 20{M}pixel 3-layer stacked {CMOS} Image Sensor with {DRAM}}, 
  year={2017},
  volume={},
  number={},
  pages={76--77},
}

@misc{graphcore_bow_ipu,
  author       = {{Graphcore}},
  title        = {Bow {IPU} Processors},
  howpublished = {\url{https://www.graphcore.ai/bow-processors}},
  year         = {2026},
}

@article{xu2018powerful,
  title={How powerful are graph neural networks?},
  author={Xu, Keyulu and Hu, Weihua and Leskovec, Jure and Jegelka, Stefanie},
  journal={arXiv preprint: 1810.00826},
  year={2018}
}

@misc{
anonymous2026efficientrefiner,
title={{EfficientRefiner}: An Efficient Refinement Method over Black-Box Optimization in Macro Placement},
author={Deng, Ji and Gao, Jun and Li, Zhao and Zhang, Ji},
year={2026},
url={https://openreview.net/forum?id=KBNvtUfjZL}
}

@INPROCEEDINGS{yu2021foundry,
  author={Yu, Douglas C. H. and Wang, Chuei-Tang and Hsia, Harry},
  booktitle=IEDM, 
  title={Foundry Perspectives on {2.5D/3D} Integration and Roadmap}, 
  year={2021},
  volume={},
  number={},
  pages={3.7.1-3.7.4}
}

@ARTICLE{kim2013study,
  author={Kim, Dae Hyun and Athikulwongse, Krit and Lim, Sung Kyu},
  journal=TVLSI, 
  title={Study of Through-Silicon-Via Impact on the {3-D} Stacked {IC} Layout}, 
  year={2013},
  volume={21},
  number={5},
  pages={862-874},
}

@INPROCEEDINGS{panth2014design,
  author={Panth, Shreepad and Samal, Sandeep and Yu, Yun Seop and Lim, Sung Kyu},
  booktitle={Proc. SOI-3D-Subthreshold Microelectronics Technology Unified Conference}, 
  title={Design challenges and solutions for ultra-high-density monolithic 3D ICs}, 
  year={2014},
  volume={},
  number={},
  pages={1-2},
}

@inproceedings{ren2026reinforcement,
  title={Reinforcement Learning for Hybrid Bonding Terminal Legalization in {3D ICs}},
  author={Ren, Wanqi and Gao, Chengrui and Shi, Yunqi and Fan, Mingzhou and Xu, Siyuan and Xue, Ke and Ding, Chenjian and Yuan, Mingxuan and Qian, Chao},
  booktitle=DATE,
  year={2026}
}

\end{document}